%% file: causalinfmultexp.tex
\documentclass[12pt]{article}
\usepackage{moreverb,url}
\RequirePackage{amsmath,amssymb}
\usepackage{animate}
\usepackage{amsthm}
\usepackage{amsmath}
\usepackage[toc,page]{appendix}
\usepackage{array}
\usepackage{bm}
\usepackage{url}
\usepackage{color}
\usepackage{float}
\usepackage{graphicx}
\usepackage{epstopdf}
\usepackage[letterpaper,portrait,margin=1in]{geometry}
\usepackage{listings}
\usepackage{mathtools}
\usepackage{natbib}
\usepackage{fancyref}
\usepackage{fancyhdr}
\usepackage[dvipsnames]{xcolor}
\usepackage{afterpage}
\usepackage{soul}

\newcommand\BibTeX{{\rmfamily B\kern-.05em \textsc{i\kern-.025em b}\kern-.08em
T\kern-.1667em\lower.7ex\hbox{E}\kern-.125emX}}

\theoremstyle{definition}

\theoremstyle{remark}
\newtheorem*{remark}{Remark}
\newtheorem{result}{Result}

\begin{document}
\def\spacingset#1{\renewcommand{\baselinestretch}
{#1}\small\normalsize} \spacingset{1}

\afterpage{\cfoot{\thepage}}

\title{Causal Inference for Multiple Treatments using Fractional Factorial Designs \thanks{\noindent{\textit{Email: } \texttt{nicole.pashley@rutgers.edu}. The authors thank Zach Branson, Kristen Hunter, Kosuke Imai, Xinran Li, Luke Miratrix, and Alice Sommer for their comments. 
They also thank Donald B. Rubin and Tirthankar Dasgupta for their insights and prior work that made this paper possible.
Additionally, they thank members of Marie-Ab\` ele Bind's research lab and Luke Miratrix's C.A.R.E.S. lab, as well as participants of STAT 300, the Harvard IQSS Workshop, and the Yale Quantitative Research Methods Workshop for their feedback on this project.
Marie-Ab\` ele Bind was supported by the John Harvard Distinguished Science Fellows Program within the FAS Division of Science of Harvard University and is supported by the Office of the Director, National Institutes of Health under Award Number DP5OD021412.
Nicole Pashley was supported by the National Science Foundation Graduate Research Fellowship under Grant No. DGE1745303. The content is solely the responsibility of the authors and does not necessarily represent the official views of the National Institutes of Health or the National Science Foundation.}}}

\author{Nicole E. Pashley\\Rutgers University \and Marie-Ab\` ele C. Bind\\
Massachusetts General Hospital}

\maketitle

\begin{abstract}
We consider the design and analysis of multi-factor experiments using fractional factorial and incomplete designs within the potential outcome framework.
These designs are particularly useful when limited resources make running a full factorial design infeasible.
We connect our design-based methods to standard regression methods.
We further motivate the usefulness of these designs in multi-factor observational studies, where certain treatment combinations may be so rare that there are no measured outcomes in the observed data corresponding to them.
Therefore, conceptualizing a hypothetical fractional factorial experiment instead of a full factorial experiment allows for appropriate analysis in those settings.
We illustrate our approach using biomedical data from the 2003-2004 cycle of the National Health and Nutrition Examination Survey to examine the effects of four common pesticides on body mass index.
\end{abstract}


\def\spacingset#1{\renewcommand{\baselinestretch}
{#1}\small\normalsize} \spacingset{1.5}
\section{Introduction}

What is the effect of exposing you to pesticide A, compared to no exposure, on your health?
What about exposing you to pesticide B? Or exposing you to both pesticides at the same time?
To answer these questions, we need to estimate the effects, including interactions, of multiple treatments.
A randomized factorial experiment uses a random assignment of all possible treatment combinations to units to estimate these different effects.
There is much interest in assessing the effects of multiple treatments, as reflected by the recent growth in the literature regarding the use of factorial designs in causal inference \citep[e.g.,][]{branson2016improving, Dasgupta2015, dong2015using, egami2017causal, espinosa2016bayesian, lu2016covariate, lu2016randomization,  lu2017randomization, mukerjee2017using, Zhao2016}.

However, a full factorial design, in which all possible combinations of treatments are randomized, may not be possible due to constraints on resources such as units or cost.
Instead, a fractional factorial or an incomplete factorial design, which uses a subset of all possible treatment combinations, may be more practical.
We layout a Neyman-style randomization-based framework for the design and analysis of these designs using potential outcomes.
Consideration of observational studies with multiple factors, in which certain treatment combinations may be missing from the data set, further motivates the usefulness of these designs.

With only a subset of all treatment combinations, saturated linear regression models have underlying assumptions resulting in estimators that are not always transparent, especially with respect to the implicit imputation of the missing potential outcomes.
We discuss regression estimates in such setting in Section~\ref{sec:part_fac} and assess how they connect to design-based estimates.
Analyses using regression models in observational studies are even trickier, as the use of regression without a careful design phase, in which one tries to uncover or approximate some underlying randomized experiment, can lead to incorrect conclusions \citep[e.g., see][]{rubin2008objective}.
Despite this difficulty, regression models with interaction terms are commonly used to estimate the effects of multiple treatments, in particular in observational studies \citep{Bobb15, Oulhote17, Patel10, Valeri17}.
For instance, \citet{Bobb15} considers a Bayesian kernel machine regression for estimating the health effects of multi-pollutant mixtures.
Therefore, exploration of design-based methods for multi-factor observational studies is an important missing piece in the literature.

The paper proceeds as follows:
Section~\ref{sec:full_fac} reviews full factorial designs within the potential outcomes framework described in \citet{Dasgupta2015}.
Section~\ref{sec:frac_fac} discusses extensions of this framework to fractional and incomplete factorial designs, expanding upon current inference results for variance and variance estimation.
Section~\ref{sec:obs_stud} examines how to embed an observational study into one of these experimental designs and illustrates our method and the challenges when working with observational data with multiple treatments with an application examining the effects of four pesticides on body mass index (BMI) using data from the National Health and Nutrition Examination Survey (NHANES), which is conducted through the Centers for Disease Control and Prevention (CDC).
Section~\ref{sec:conc} concludes.

Throughout, we will use a running example of an experiment examining the impact of three pesticides (which we call pesticides A, B, and C) on weight gain in zebrafish.
Our data example provides an illustration of the way we foresee these designs being useful when experimentation is not possible, such as when discussing the impact of pesticides on humans.

\section{Set up}\label{sec:full_fac}
\subsection{Notation and factorial estimands}\label{subsec:est_fac}

We closely follow the potential outcomes \citep{neyman_1923, rubin_1974} framework for $2^K$ factorial designs in \citet{Dasgupta2015}. We focus on designs with multiple two-level factors assigned in combination; that is, there are a number of distinct treatments (e.g., medications), each having two levels (e.g., receiving a certain medication or not).
Let there be $K$ two-level factors, which creates $2^K = J$ total possible treatment combinations.
The $j^{th}$ treatment combination is denoted $\bm{z}_j$.
See Table~\ref{tab:fac_design_example} for an example with $K=3$, which illustrates the notation. For consistency, we list treatment combinations in lexicographic order, the standard ordering.
Let $z_{j,k} \in \{-1, +1\}$ be the level of the $k^{th}$ factor in the $j^{th}$ treatment combination, so $\bm{z}_j = (z_{j,1},...,z_{j,K})$.
Let there be $n$ units in the sample.
We assume the Stable Unit Treatment Value Assumption (SUTVA): there is no interference and no different forms of each treatment level \citep[][]{rubin_1980}.
Then the potential outcome for unit $i$ under treatment combination $\bm{z}_j$ can be written as $Y_i(\bm{z}_j)$ and the sample average potential outcome under treatment $\bm{z}_j$ as $\bar{Y}(\bm{z}_j) = \sum_{i=1}^nY_i(\bm{z}_j)/n$.
$\bar{\bm{Y}} = (\bar{Y}(\bm{z}_1), \bar{Y}(\bm{z}_2), \hdots, \bar{Y}(\bm{z}_{J}))$ is the vector of mean potential outcomes for all $2^K$ treatment combinations.

\def\spacingset#1{\renewcommand{\baselinestretch}
{#1}\small\normalsize} \spacingset{1}
\begin{table}[ht]
\centering
\begin{tabular}{ccccc}
\hline
Treatment & Factor $1$ & Factor $2$ & Factor $3$ & Outcomes\\
\hline
$\bm{z}_1$ & -1 & -1 & -1 & $\bar{Y}(\bm{z}_1)$\\
$\bm{z}_2$ & -1 & -1 & +1 & $\bar{Y}(\bm{z}_2)$\\
$\bm{z}_3$ & -1 & +1 & -1 & $\bar{Y}(\bm{z}_3)$\\
$\bm{z}_4$ & -1 & +1 & +1 & $\bar{Y}(\bm{z}_4)$\\
$\bm{z}_5$ & +1 & -1 & -1 & $\bar{Y}(\bm{z}_5)$\\
$\bm{z}_6$ & +1 & -1 & +1 & $\bar{Y}(\bm{z}_6)$\\
$\bm{z}_7$ & +1 & +1 & -1 & $\bar{Y}(\bm{z}_7)$\\
$\bm{z}_8$ & +1 & +1 & +1 & $\bar{Y}(\bm{z}_8)$\\
\hline
&$\bm{g}_1$&$\bm{g}_2$&$\bm{g}_3$&$\bm{\bar{Y}}$ \rule{0pt}{0.4cm}\\
\hline
\end{tabular}
\caption{Example of a $2^3$ factorial design.}
\label{tab:fac_design_example}
\end{table} 
\def\spacingset#1{\renewcommand{\baselinestretch}
{#1}\small\normalsize} \spacingset{1.5}

Using the framework from \citet{Dasgupta2015}, denote the contrast column $j$ in the design matrix by $\bm{g}_{j}$, as illustrated in Table~\ref{tab:fac_design_example}.
Following that paper, we can also define the contrast vector for the two-factor interaction between factors $k$ and $k'$ as
$\bm{g}_{k,k'} = \bm{g}_{k}\circ \bm{g}_{k'}$,
where $\circ$ indicates the Hadamard (element-wise) product.
Similarly, the contrast vector for three-factor interactions is
$\bm{g}_{k,h, i} = \bm{g}_{k}\circ \bm{g}_{h} \circ \bm{g}_{i}$,
and all higher order interaction contrast vectors can be calculated analogously.

Continuing to follow \citet{Dasgupta2015}, define the finite population main causal effect for factor $k$ and the finite population interaction effect for  factors $k$ and $k'$ as
\[\tau(k) = \frac{1}{2^{K-1}}\bm{g}^T_k\bar{\bm{Y}} \ \ \text{and} \ \ \tau(k,k') = \frac{1}{2^{K-1}}\bm{g}^T_{k, k'}\bar{\bm{Y}},\]
respectively, where by finite population we mean that we are only interested in inference for the units we have in the experiment. 
Higher-level interaction terms are defined analogously.
We can similarly define the individual-level effects as
\[\tau_i(k)=\frac{1}{2^{K-1}}\bm{g}^T_k\bm{Y}_i  \ \  \text{and}  \ \ \tau_i(k, k') = \frac{1}{2^{K-1}} \bm{g}^T_{k, k'}\bm{Y}_i,\]
where $\bm{Y}_i = (Y_i(\bm{z}_1), \hdots, Y_i(\bm{z}_J))^T$.
We also define the average potential outcome across treatments as $\tau(0)  = 2^{-K}\bm{g}^T_{0}\bar{\bm{Y}}$,
where $\bm{g}_{0}$ is a vector of length $2^K$ of all $+1$'s.

Consider our example of estimating the effects of pesticides A, B, and C on weight gain in zebrafish.
We take the three pesticides to be three factors, each with two levels: applying the pesticide (+1) or not (-1).
In Table~\ref{tab:fac_design_example}, zebrafish assigned to $\bm{z}_6$ would be those that get pesticides A and C, but not B.
The main effect of pesticide A ($\tau(1)$) is the average effect of receiving pesticide A, compared to not receiving it, uniformly averaged over all combinations of receiving or not receiving pesticides B and C.
The two-factor interaction between pesticides A and B, would be the difference in the effect of receiving pesticide A, compared to not receiving it, when pesticide B is applied versus when it is not, averaging uniformly over receiving or not receiving pesticide C.

Consider a full factorial model matrix, $\bm{G}$, that includes the mean and interactions and whose rows are comprised of $\bm{g}_k^T$, where $k \in \{0; 1; ...; K; (1,2); ...;(1,2, \dots, K)\}$.
Based on the definition of our estimands in Section~\ref{subsec:est_fac}, the matrix $\bm{G}$ relates the mean potential outcomes and the factorial effects as follows: Let $\bm{G}$ be a matrix with the $\bm{g}$ vectors as rows,  
\[\bm{G} = [\bm{g}_{0}, \bm{g}_{1}, \dots, \bm{g}_{K}, \bm{g}_{1,2}, \dots, \bm{g}_{1, \dots, K}]^T\]
and $\bm{\tau}$ be the column vector of all factorial effects, 
\[\bm{\tau} = [2\tau(0), \tau(1), \dots, \tau(K), \tau(1,2), \dots, \tau(K-1, K), \dots, \tau(1,2,\dots,K)]^T.\]
Then, 
\[2^{-(K-1)}\bm{G}\bm{Y} = \bm{\tau}.\]

\noindent Due to orthogonality, $\bm{G}^{-1}=\frac{1}{2^K}\bm{G}^T$, as argued by \citet{Dasgupta2015} in the context of imputing potential outcomes under Fisher's sharp null hypothesis.
The mean potential outcomes can be rewritten in terms of the factorial effects as
${\bar{\bm{Y}}=2^{-1}\bm{G}^T\bm{\tau}}$.
Let the $j^{th}$ row of $\bm{G}^T$ be denoted by $\bm{h}_j$.
The first entry of $\bm{h}_j$ is $+1$, corresponding to the mean, the next $K$ entries  are equal to the entries of $\bm{z}_{j}$, and the remaining entries correspond to interactions, with the order given by the order of the rows of $\bm{G}$.
In the $2^3$ factorial design shown in Table~\ref{tab:fac_design_example}, $\bm{h}_1 = (+1, -1, -1, -1, +1, +1, +1, -1)$.
That is, it is the entries of the first row of Table 1, prepended with an entry for $2\tau(0)$ and appended with entries for the interactions.
We have $\bar{Y}(\bm{z}_j) = 2^{-1}\bm{h}_j\bm{\tau}$.
We will use this representation in Section~\ref{sec:part_fac} when considering causal estimands in settings lacking units for some treatment combinations.

\subsection{Estimation and inference in a $2^K$ factorial design}
We now review inference for $2^K$ factorial designs.
Let $W_i(\bm{z}_j) = 1$ if unit $i$ received treatment $\bm{z}_j$ and $W_i(\bm{z}_j) = 0$ otherwise.
In a full $2^K$ factorial design, under complete randomization, all $2^K$ factors are assigned to units completely at random (possibly unbalanced).
That is, there is a fixed number, $n_{j}\geq 2$, of units randomly assigned to treatment combination $\bm{z}_j$.
Let $\bm{W}$ be the $n \times 2^K$ vector of $W_i(\bm{z}_j)$ for all $i$ and $\bm{z}_j$.
Let $\bm{w}$ be some $n \times 2^{K}$ matrix with entries in the set $\{0,1\}$.
This design imposes 
\[P(\bm{W} = \bm{w})= 
\prod_{j=1}^{J}n_j!/n! \]
if $\sum_{j=1}^{J}w_i(\bm{z}_j) = 1$  for all  $i$ and $\sum_{i=1}^nw_i(\bm{z}_j) = n_j$  for all  $\bm{z}_j$, and $P(\bm{W} = \bm{w}) = 0$ otherwise.
The observed potential outcome for unit $i$ is $Y_i^{obs} = \sum_{j=1}^JW_i(\bm{z}_j)Y_i(\bm{z}_j)$.
An unbiased estimator of the average potential outcome for treatment $\bm{z}_j$ is
\[\bar{Y}^{obs}(\bm{z}_j) = \frac{1}{n_{j}}\sum_{i=1}^nW_i(\bm{z}_j)Y_i(\bm{z}_j) = \frac{1}{n_{j}}\sum_{i: W_i(\bm{z}_j)=1}^nY_i^{obs}.\] 
Denote $\bm{\bar{Y}}^{obs}= (\bar{Y}^{obs}(\bm{z}_1), \bar{Y}^{obs}(\bm{z}_2), \hdots, \bar{Y}^{obs}(\bm{z}_{J}))$ as the vector of observed mean potential outcomes for all $2^K$ treatment combinations.

Unbiased estimators for $\tau(k)$ and $\tau(k, k')$ are defined as
\[\widehat{\tau}(k) = \frac{1}{2^{K-1}}\bm{g}^T_k\bar{\bm{Y}}^{obs}  \ \ \text{and} \ \  \widehat{\tau}(k, k') = \frac{1}{2^{K-1}}\bm{g}^T_{k, k'}\bar{\bm{Y}}^{obs},\]
respectively.
Estimators for higher-level interaction terms and $\tau(0)$ are defined analogously.

The variance of the factorial effect estimators under an unbalanced design is given in \citet{lu2016randomization} \citep[extended from balanced case in]
[]{Dasgupta2015}, as follows:
\begin{align}
\text{Var}\left(\widehat{\tau}(k)\right) = \frac{1}{2^{2(K-1)}}\sum_{j=1}^J\frac{1}{n_j}S^2(\bm{z}_j) -\frac{1}{n}S_{k}^2, \label{eq:varfullfac}
\end{align}
where
\[S^2(\bm{z}_j) = \frac{1}{n-1}\sum_{i=1}^n\left(Y_i(\bm{z}_j) - \bar{Y}(\bm{z}_j)\right)^2 \quad \text{and} \quad S^2_k = \frac{1}{n-1}\sum_{i=1}^n\left(\tau_i(k) -\tau(k)\right)^2.\]

An expression for the covariance between two factorial effect estimators for unbalanced factorial designs is given by \citet{lu2016randomization} \citep[extended from][]{Dasgupta2015} as
\begin{align}
\text{Cov}\left(\hat{\tau}(k), \hat{\tau}(k')\right)= \frac{1}{2^{2(K-1)}}\left[\sum_{j: g_{kj}=g_{k'j}}\frac{1}{n_j}S^2(\bm{z}_j)-\sum_{j: g_{kj}\neq g_{k'j}}\frac{1}{n_j}S^2(\bm{z}_j)\right]-\frac{1}{n}S_{k,k'}^2,\label{eq:covfullfac}
\end{align}
where $S^2_{k,k'}=\sum_{i=1}^n\left(\tau_i(k)-\tau(k)\right)\left(\tau_i(k')-\tau(k')\right)/(n-1)$.

Further discussion of Neymanian and Fisherian inference for full factorial designs is reviewed in Supplementary Material~\ref{sec:fac_stat_inf}.

\section{Alternate designs to full factorial: Fractional and incomplete}\label{sec:frac_fac}

\subsection{Fractional factorial designs}
There are situations in which a full $2^{K}$ factorial design experiment cannot be conducted or is suboptimal.
Limited resources may mean there are an insufficient number of units to randomly assign to each of the $J$ ($= 2^{K}$) treatment groups. Or the full factorial design may not be the most efficient allocation of resources; for example, if the experimenter believes that higher-order interactions are negligible \citep{WuHamada}.
Instead, an experimenter might implement a $2^{K-p} = J'$ fractional factorial design in which only $J'$ of the total $J$ treatment combinations are used.
Here we will give a brief overview of this design, but we recommend \citet{WuHamada} for a more detailed review.
We follow notation in \citet{Dasgupta2015} and \citet{stat240dontd}.

To create a $2^{K-p}$ design, we can write out a full factorial design for the first $K-p$ factors, and fill in the $p$ columns for the remaining factors using multiplicative combinations of subsets of the previous contrast columns.
We use a generator to choose the factors whose treatment levels we multiply together to get the treatment levels for the other factors \citep{WuHamada}.
For example, to create a $2^{3-1}$ design using the generator ``$3=12$'', we write out a $2^2$ design for factors 1 and 2, as in column 2 and 3 of Table~\ref{tab:frac_fac_design_example}.
We generate the third column, corresponding to treatment levels for factor 3, by multiplying together the contrast vectors for factors 1 and 2, as shown in Table~\ref{tab:frac_fac_design_example}.
Which two factors are used initially is irrelevant in this case because of symmetry.
Because $3=12$, we also have $1=23$ and $2=13$.
However, this may not hold in general and one should choose which factors to use based on the final aliasing structure, which tells us which effects are confounded.

\def\spacingset#1{\renewcommand{\baselinestretch}
{#1}\small\normalsize} \spacingset{1}
\begin{table}[ht]
\centering
\begin{tabular}{ccccc}
\hline
Treatment & Factor $1$ & Factor $2$ & Factor $3$ & Outcomes\\
\hline
$\bm{z}_1^*$ & -1 & -1 & +1 & $\bar{Y}(\bm{z}_1^*)$\\
$\bm{z}_2^*$ & -1 & +1 & -1 & $\bar{Y}(\bm{z}_2^*)$\\
$\bm{z}_3^*$ & +1 & -1 & -1 & $\bar{Y}(\bm{z}_3^*)$\\
$\bm{z}_4^*$ & +1 & +1 & +1 & $\bar{Y}(\bm{z}_4^*)$\\
\hline
&$\bm{g}_1^*$&$\bm{g}_2^*$&$\bm{g}_3^*$&$\bar{\bm{Y}}_*$ \rule{0pt}{0.4cm}\\
\hline
\end{tabular}
\caption{Example of a $2^{3-1}$ factorial design.}
\label{tab:frac_fac_design_example}
\end{table}
\def\spacingset#1{\renewcommand{\baselinestretch}
{#1}\small\normalsize} \spacingset{1.5}

Under this design, the contrast columns, $\bm{g}_k^*$, are now shortened versions or subsets of the contrast columns of a full $2^3$ factorial design, $\bm{g}_k$.
The treatment combinations in the fractional design, $\bm{z}^*_j$, are a subset of the full set of treatment combinations, $\bm{z}_j$, for the $2^3$ design.
Referring to Table~\ref{tab:frac_fac_design_example}, $\bm{g}_{3}^* = \bm{g}_1^* \circ \bm{g}_2^*$.
The generator $3=12$ indicates that the main effect of factor 3 is aliased with the two-factor interaction 12, which means that we cannot distinguish these effects -- they are confounded and combined in our estimators.
The full aliasing in this design is as follows: $I = 123$, $3=12$, $2=13$, $1=23$, where $I$ corresponds to a vector of all +1's ($\bm{g}_0$).
The relation $I=123$ is called the defining relation, which characterizes the aliasing and how to generate the rest of the columns.
The main effects, as defined in Section~\ref{subsec:est_fac}, are aliased with the two-factor interactions.
If the two-factor interactions are negligible, the 2$^{3-1}$ design is a parsimonious design to estimate the main effects and we can construct unbiased estimators for the main effects (see Section~\ref{subsec:frac_fac_est}). The resulting design is also orthogonal (i.e., all pairs of columns are orthogonal) and balanced in the sense that each $\bm{g}^*_{k}$ has an equal number +1's and -1's \citep{WuHamada}.
These properties simplify the aliasing structure.

We typically choose the generator or defining relation based on the maximum resolution criterion for the design.
Resolution is defined as the word length (i.e., the number of factors) in the shortest word of the defining relation \citep[see][for more details]{WuHamada} and indicates the aliasing structure and which levels of effects the main effects are aliased.
In our example of a $2^{3-1}$ design, we only have one defining relation, $I=123$, the word is length 3, and
main effects are aliased with two-factor interactions.
Aliasing some main effects with other main effects is an alternate aliasing structure.
The effect hierarchy principle assumes that lower-order effects have larger magnitude than higher-order interaction effects \citep{WuHamada}.
Therefore, one generally chooses a fractional design where main effects and some other lower-order interaction effects are aliased with higher-order terms, i.e., the main effects and two-factor interactions are clear. This assumption goes along with the assumption of effect sparsity (the number of important effects is small), as a justification of the use of fractional designs over the full factorial \citep{WuHamada}.
It is possible that there may be multiple defining relations that lead to the same resolution.
In addition to the maximum resolution criterion, it is standard to consider minimum aberration, which says we want the design with the smallest number of short words in its defining relations.

\subsubsection{Estimators}\label{subsec:frac_fac_est}

In this section we review estimators for the fractional design, which are similar to the full factorial case, and follow the framework laid out in \citet{Dasgupta2015} and \citet{stat240dontd}.
We assume units are assigned using complete randomization to the $J'$ treatments.
In other words, if $\bm{W}^*$ is the $n \times 2^{K-p}$ matrix of $W_i(\bm{z}_j^*)$ for all $i$ and $\bm{z}_j$, and $\bm{w}^*$ is some $n \times 2^{K-p}$ matrix with entries in the set $\{0,1\}$, then
\[P(\bm{W}^* = \bm{w}^*)= 
\prod_{j=1}^{J'}n_j^*!/n! \]
if $\sum_{j =1}^{J'}w_i(\bm{z}_j^*) = 1$  for all  $i$ and $\sum_{i=1}^nw_i(\bm{z}_j^*) = n_j^*$  for all  $\bm{z}_j^*$, and $P(\bm{W}^* = \bm{w}^*) = 0$ otherwise.
Randomization is the same as in the full factorial case, but with fewer treatment groups.
All of the following inference results are derived
under complete randomization.

The estimator for $\tau(k) $ is defined as
\[\widehat{\tau}^*(k) = \frac{1}{2^{K-p-1}}\bm{g}_k^{*T}\bar{\bm{Y}}^{obs}_*.\]

\noindent The estimator for $\tau(k, k')$ is defined as
\[\widehat{\tau}^*(k, k') = \frac{1}{2^{K-p-1}}\bm{g}_{k, k'}^{*T}\bar{\bm{Y}}^{obs}_*.\]

\noindent These estimators are no longer unbiased.
Let $\mathcal{A}[k]$ be the set of all effects aliased with factorial effect $k$ as well as $k$ itself.
The number of factors aliased with factor $k$ is $2^p-1$ \citep[][Chapter 8]{montgomery2017design}.
So, $\mathcal{A}[k]$ has $2^p$ elements.
Factor $k$ may be aliased with the negative of a main effect or interaction.
Let $A_{k,j}$ be the indicator for whether factor $j$ is negatively aliased with factor $k$ ($A_{k,j} = 1$) or positively aliased ($A_{k,j} = 0$).

\begin{result}\label{result:frac_bias}
The expectation of the factorial estimator is
\begin{align*}
E\left[\widehat{\tau}^*(k)\right] = \frac{1}{2^{K-p-1}}\bm{g}_k^{*T}\bar{\bm{Y}}_*
=\sum_{j \in \mathcal{A}[k]} (-1)^{A_{k,j}}\tau(j)
=\tau(k) + \sum_{j \in \mathcal{A}[k]\backslash \{k\}} (-1)^{A_{k,j}}\tau(j).
\end{align*}
\end{result}
Result~\ref{result:frac_bias} can be shown using the orthogonality of the $\bm{g}_k$ vectors.
Hence, by aliasing these effects, they get combined in our estimator.

Returning to our running example, let us say we used the design in Table~\ref{tab:frac_fac_design_example}.
In that design, the main effect of pesticide A is aliased with the interaction between pesticide B and C.
The expectation of the main effect for pesticide A would be $E\left[\widehat{\tau}^*(1)\right] = \tau(1) + \tau(23)$.
If the the interaction effect between pesticides B and C is small, our estimator for the main effect of pesticide A will be close to unbiased.
Otherwise, our estimate of the effect of pesticide A will be biased by the size of the interaction.
As an example, if the main effect of pesticide A is $\tau(1) = -2$ and the interaction between pesticides B and C is $\tau(23) = 2$, the expectation of our estimator will be 0 even though the true average effect of pesticide A is nonzero.

More generally, the estimator for $\tau(k) $ is unbiased if the effects aliased with factorial effect $k$ are zero, and will be close to unbiased if the aliased effects are negligible, which may be justified by the effect hierarchy principle.
That is, we expect higher-order effects to be smaller in magnitude.
When aliasing occurs such that main effects are aliased with low-order interaction terms, such as two-factor interactions, this assumption may be unrealistic.
However, if we have a large number of factors and are only aliasing main effects with higher-order effects, this may be reasonable.
For example, in a $2^{6-1}$ fractional design, main effects can be aliased with five-factor interactions and two-factor interactions can be aliased with four-factor interactions, where both higher-order interactions may be assumed to be small.

We extend the variance and variance estimator expressions from \citet{Dasgupta2015} (see Section~\ref{sec:full_fac}) to this setting. 
Recall $J' = 2^{K-p}$ and define $\tilde{\tau}_i(k) = \tau_i(k) + \sum_{j \in \mathcal{A}[k]\backslash \{k\}} (-1)^{A_{k,j}}\tau_i(j)$ and $\tilde{\tau}(k) = \frac{1}{N}\sum_{i=1}^N\tilde{\tau}_i(k) =  E[\widehat{\tau}^*(k)]$.
Further define the analog of $S^2_k$,
\[\tilde{S}^2_k = \frac{1}{n-1}\sum_{i=1}^n\left(\tilde{\tau}_i(k) - \tilde{\tau}(k)\right)^2,\]
which is the variation in our newly defined aliased effects, $\tilde{\tau}_i(k)$. Let $n^*_j$ be the number of units assigned to treatment combination $\bm{z}^*_j$.
\begin{result}
The variance of the estimator $\hat{\tau}^*(k)$ is
\begin{align}
\text{Var}\left(\hat{\tau}^*(k)\right) &=\frac{1}{2^{2(K-p-1)}}\sum_{j=1}^{J'}\frac{1}{n^*_j}S^2(\bm{z}^*_j) -\frac{1}{n}\tilde{S}_k^2. \label{eq:varfracfac}
\end{align}
\end{result}

We can compare the variance under a factorial design to one under a fractional factorial design easily in a simplified setting.
Take sample size to be $n$ for both designs, and both designs to be balanced ($n_j = n/J$, $n_j^* = n/J'$).
Further, assume we have additive effects (no effect heterogeneity), which implies $S^2(\bm{z}_j) = S^2$ for all $\bm{z}_j$ and $S_k^2= \tilde{S}_k^2 = 0$.
In this setting, the variance of the estimator for $\widehat{\tau}(k)$ for the full factorial design will be
\[\text{Var}\left(\widehat{\tau}(k)\right) = \frac{1}{2^{2(K-1)}}\frac{J^2}{n}S^2 = \frac{4}{n}S^2\]
and for the fractional factorial design will be
\[\text{Var}\left(\widehat{\tau}^*(k)\right) = \frac{1}{2^{2(K-p-1)}}\frac{(J')^2}{n}S^2 = \frac{4}{n}S^2.\]
The variance of the estimator under the full factorial design will be equal to that under a fractional factorial design.
As the fractional factorial design has a bias when the interaction effects being aliased are nonzero, the full factorial is preferred when possible.
The general case without additive effects and a balanced design is more complicated, especially if the treatments included in the fractional factorial design happen to be those that produce less variable outcomes.
However, this would be difficult to predict before running the experiment, and therefore a full factorial is preferred in general.
Unbalanced designs will similarly change this result, potentially increasing or decreasing the weight on more variable groups in the two designs.
A full factorial design may not always be feasible or practical, in which case inducing some (hopefully negligible) bias by using a fractional factorial design may be acceptable.

We can also obtain the covariance  between two fractional factorial effect estimators.
We define an altered version of $S^2_{k,k'}$:
\begin{align}
\tilde{S}^2_{k,k'}&=\frac{1}{n-1}\sum_{i=1}^n\left(\tilde{\tau}_i(k)-\tilde{\tau}(k)\right)\left(\tilde{\tau}_i(k')-\tilde{\tau}(k')\right).
\end{align}
\begin{result}
The covariance of $\hat{\tau}^*(k)$ and $\hat{\tau}^*(k')$ is
\begin{align}
\text{Cov}\left(\hat{\tau}^*(k), \hat{\tau}^*(k')\right)= \frac{1}{2^{2(K-p-1)}}\left[\sum_{j: g_{kj}^*=g_{k'j}^*}\frac{1}{n^*_j}S^2(\bm{z}^*_j) -\sum_{j: g_{kj}^*\neq g_{k'j}^*}\frac{1}{n^*_j}S^2(\bm{z}^*_j) \right]-\frac{1}{n}\tilde{S}_{k,k'}^2.
\end{align}
\end{result}
These variance and covariance expressions result from an application of Theorem 3 in \citet{li2017general}, as explained in Appendix~\ref{apped:subsecvarfracfac}. 
The variance expressions for fractional factorial designs are similar to the full factorial case, but are defined over aliased effects.
In particular, the treatment effect heterogeneity term has the interpretation of the heterogeneity of a linear combination of aliased effects.

\subsubsection{Statistical inference}\label{subsec:frac_fac_stat_anal}

It is straightforward to extend the analysis for factorial designs in \citet{Dasgupta2015} and reviewed in Section~\ref{sec:full_fac} to fractional factorial designs. We can use a conservative Neyman-style variance estimator, similar to the full factorial setting (see Supplementary Material~\ref{sec:fac_stat_inf}):
\begin{align}
\widehat{\text{Var}}\left(\hat{\tau}^*(k)\right)
&= \frac{1}{2^{2(K-p-1)}}\sum_{j=1}^{J'}\frac{1}{n_j^*}s^2(\bm{z}^*_j).
\end{align}
\begin{result}
The expectation of the fractional factorial variance estimator is
\[E\left[\widehat{\text{Var}}\left(\hat{\tau}^*(k)\right)\right] = \frac{1}{2^{2(K-p-1)}}\sum_{j=1}^{J'}\frac{1}{n_j^*}S^2(\bm{z}^*_j).\]
\end{result}
Thus $\widehat{\text{Var}}\left(\hat{\tau}^*(k)\right)$ is a conservative estimator and unbiased if and only if $\tilde{S}_k^2 = 0$, which would occur if the $\tilde{\tau}_i(k)$ are constant. This condition holds when all effects aliased with factor $k$ are constant additive effects.

We can build confidence regions and confidence intervals analogously to what was reviewed in Section~\ref{sec:fac_stat_inf}.
To be more rigorous in this section, we show how to use the results of \citet{li2017general} to build asymptotic Wald-type confidence regions.
There is a new matrix $\bm{G}^*$ that relates the treatment combinations in the fractional experiment to the unique treatment effect estimands and estimators defined for this experiment.
For the $2^{3-1}$ example, recalling the aliasing structure which gives $\bm{g}^{*}_{0} = \bm{g}^{*}_{123}$ and $\tilde{\tau}(2) = \tilde{\tau}(13)$, we have

\def\spacingset#1{\renewcommand{\baselinestretch}
{#1}\small\normalsize} \spacingset{1}
\begin{align*}
2^{-(K-p-1)}\underbrace{\begin{bmatrix}
\bm{g}^{*T}_{0}\\
\bm{g}^{*T}_1\\
\bm{g}^{*T}_2\\
\bm{g}^{*T}_3\\
\end{bmatrix}}_{\bm{G}^*}
\underbrace{\begin{bmatrix}
\bar{Y}(\bm{z}^*_1)\\
\bar{Y}(\bm{z}^*_2)\\
\bar{Y}(\bm{z}^*_3)\\
\bar{Y}(\bm{z}^*_4)\\
\end{bmatrix}}_{\bm{\bar{Y}}^*} = 
\underbrace{\begin{bmatrix}
2\tilde{\tau}(0)\\
\tilde{\tau}(1)\\
\tilde{\tau}(2)\\
\tilde{\tau}(3)\\
\end{bmatrix}}_{\tilde{\bm{\tau}}}.
\end{align*}
\def\spacingset#1{\renewcommand{\baselinestretch}
{#1}\small\normalsize} \spacingset{1.5}
Define $\hat{\bm{\tau}}^{*}$ as the $2^{K-p}$ vector of unique treatment effect estimators that corresponds to estimands in $\tilde{\bm{\tau}}$.
\begin{result}\label{result:clt}
Assume that all $S^2(\bm{z}^*_j)$ and $\tilde{S}_{k,k'}^2$ have limiting values, that the $n^*_j/n$ have positive limiting values, and  $\text{max}_{1 \leq j \leq J'}\text{max}_{1 \leq i \leq n}\left(Y_i(z^*_j) - \bar{Y}(z^*_j)\right)^2/n \to 0$.
According to Theorem 5 of \citet{li2017general}, $n\text{var}(\hat{\bm{\tau}}^{*})$ has a limiting value, which, following their notation, we denote by $\bm{V}$ and
\[\sqrt{n}\left(\hat{\bm{\tau}}^{*} - \tilde{\bm{\tau}}\right) \xrightarrow{d} \text{N}(\bm{0}, \bm{V}).\]
\end{result}
\begin{result}
Further, let $\hat{\bm{V}} = \sum_{j=1}^{J'}n^*_j\bm{G}^{*}_js^2(\bm{z}^*_j)\bm{G}^{*T}_j$, where $\bm{G}^*_j$ is the $j$th column of $\bm{G}^*$.
In addition to the assumptions required for Result~\ref{result:clt}, we require that the limit of $\sum_{j=1}^{J'}n^*_j\bm{G}^{*}_jS^2(\bm{z}^*_j)\bm{G}^{*T}_j$ is nonsingular.
Under Proposition 3 of \citet{li2017general} and following their notation, the Wald-type confidence region,
\[\{\bm{\mu}: \left(\hat{\bm{\tau}}^{*} - \bm{\mu}\right)^T\hat{\bm{V}}\left(\hat{\bm{\tau}}^{*} - \bm{\mu}\right) \leq q_{J', 1-\alpha}\},\]
where $q_{J', 1-\alpha}$ corresponds to the $1-\alpha$ quantile of the $\chi^2_{J'}$ distribution, has at least $1-\alpha$ asymptotic coverage.
\end{result}
\noindent We assume that the number of treatments is constant as $n \to \infty$, but \citet{li2017general} discuss the case where the number of treatment combinations grows as well.

As in the full factorial setting (see \citet{lu2016randomization}, \citet{zhao2021regression}, and Supplementary Material~\ref{sec:fac_stat_inf}), linear regression on the full vector of observed outcomes, using a saturated model with factor levels coded as -1 and +1, yields the same point estimates (divided by 2) and the HC2 variance estimator yields the same variance estimate (divided by 4) as the Neymanian estimates.
For proof, see Appendix~\ref{append:linreg}.
\citet{zhao2021regression} show that weighted-least squares can also be used to obtain similar results in an unsaturated regression in the full factorial setting.

\subsection{Incomplete factorial designs}\label{sec:part_fac}
In this section, we discuss an alternative experimental design called the incomplete factorial design \citep{byar1993incomplete}, which we consider (novelly, to our knowledge) from the potential outcome and design-based perspective.
This design uses a subset of data from a full factorial design but a different subset than the fractional design.
In particular, the estimators we discuss will not use all of the available data.
We can therefore consider each estimator to be associated with a particular ``design'' that corresponds to an experiment that randomizes units only to treatment combinations used in the estimator.
 Different estimators may give non-zero weight to different treatment combinations, and so the designs may be estimand-specific.
Due to this inherent linking between the design and the estimators, we discuss the design and corresponding estimators together in the next section.
\subsubsection{Design and estimators}

Consider if we ran an experiment with $K$ binary treatments where, whether due to human error or lack of resources, some of treatment combinations from the full $2^K$ factorial design were not included in the experiment.
We may have a factorial structure as in Table~\ref{tab:partial_fac_example}, but no outcome measurements for treatment $\bm{z}_7$.
How should we analyze this?
\def\spacingset#1{\renewcommand{\baselinestretch}
{#1}\small\normalsize} \spacingset{1}
\begin{table}[ht]
\centering
\begin{tabular}{ccccc}
\hline
Treatment & $1$ & $2$ & $3$ & Observed Outcomes\\
\hline
$\bm{z}_1$ & -1 & -1 & -1  & $\bar{Y}^{obs}(\bm{z}_1)$\\
$\bm{z}_2$ & -1 & -1 & +1  & $\bar{Y}^{obs}(\bm{z}_2)$\\
\color{blue}$\bm{z}_3$ & -1 & +1 & -1  & $\bar{Y}^{obs}(\bm{z}_3)$\\
$\bm{z}_4$ & -1 & +1 & +1  & $\bar{Y}^{obs}(\bm{z}_4)$\\
$\bm{z}_5$ & +1 & -1 & -1  & $\bar{Y}^{obs}(\bm{z}_5)$\\
$\bm{z}_6$ & +1 & -1 & +1  & $\bar{Y}^{obs}(\bm{z}_6)$\\
\color{red}$\bm{z}_7$ & +1 & +1 & -1  & ?\\
$\bm{z}_8$ & +1 & +1 & +1  & $\bar{Y}^{obs}(\bm{z}_8)$\\
\hline
&$\bm{g}_1$&$\bm{g}_2$&$\bm{g}_3$&$\bar{\bm{Y}}^{obs}$ \rule{0pt}{0.4cm} \\
\hline
\end{tabular}
\caption{Example of a $2^{3}$ factorial design with no observations for one treatment combination.}
\label{tab:partial_fac_example}
\end{table}
\def\spacingset#1{\renewcommand{\baselinestretch}
{#1}\small\normalsize} \spacingset{1.5}

Focus on estimation of $\tau(1)$ first.
We could reduce the data to a fractional factorial design, which requires removing multiple treatment groups.
If we recreate a fractional factorial design that aliases the main effects with the two-way interactions, we estimate $\tau(1) + \tau(23)$, which involves using outcome data from units assigned to treatment combinations $\bm{z}_2$, $\bm{z}_3$, $\bm{z}_5$, and $\bm{z}_8$, but not from units assigned to treatment combinations $\bm{z}_1$, $\bm{z}_4$, and $\bm{z}_6$.

Instead, we might consider building an estimator using all of the treatment combinations except for $\bm{z}_3$, which has the same levels for factors 2 and 3 but the opposite level for factor 1 as combination $\bm{z}_7$.
Thus, in some sense, removing $\bm{z}_3$ ``balances'' the remaining treatment combinations and is essentially the ``na\"ive estimator'' discussed in \citet{byar1993incomplete}.
This strategy creates a different hypothetical experimental design with a different aliasing structure.
In this case, we would be estimating
\begin{align*}
\dot{\tau}(1)&=\frac{1}{3}\left[\bar{Y}(\bm{z}_5) + \bar{Y}(\bm{z}_6) + \bar{Y}(\bm{z}_8)\right]-\frac{1}{3}\left[\bar{Y}(\bm{z}_1) + \bar{Y}(\bm{z}_2) + \bar{Y}(\bm{z}_4)\right]\\
&= \frac{1}{3}\left[\tau(1|F_2=-1, F_3=-1)+\tau(1|F_2=-1, F_3=+1)+\tau(1|F_2=+1, F_3=+1)\right],
\end{align*}
where we let $F_k$ denote the level of the $k^{th}$ factor so that ${\tau(k|F_j=x,F_i=y)}$ is the main effect of factor $k$ conditional on level $x$ of factor $j$ and level $y$ of factor $i$, as in \citet{Dasgupta2015}.
In words, we estimate the average of the conditional effects of factor 1 given the combinations (-1,-1), (-1,+1), and (+1,+1) for factors 2 and 3.

To find the aliasing structure, we can refer to the matrix $\bm{G}^T$ in Section~\ref{subsec:est_fac} to rewrite the estimand above as
\begin{align*}
\dot{\tau}(1)&=\frac{1}{3}\frac{1}{2}\left(\bm{h}_8 +\bm{h}_6+\bm{h}_5 - \bm{h}_4 - \bm{h}_2 - \bm{h}_1\right)\bm{\tau}\\
&=\tau(1)+\frac{1}{3}\left[-\tau(13)+\tau(23)+\tau(123)\right].
\end{align*}
We have partially aliased the main effect for factor 1 with the two-factor interactions for factors 1 and 3 and factors 2 and 3, as well as the three-factor interaction, all divided by three.
This is partial aliasing because the factors are neither fully aliased nor completely clear of each other \citep[][Chapter 7]{WuHamada}.
Return to our pesticide example.
If $\tau(13) = 2$, $\tau(23) = 2.5$  and $\tau(123) = 1$, the main effect of pesticide A will be positively biased by $1/2$.
Whether this is preferable to aliasing the main effect with just the two-factor interaction between factors 2 and 3, as in the fractional design, depends on context.
If we know that the three-factor interaction is negligible, (or at least smaller than the two-factor interactions, as in the numerical example just given) this new estimator may have lower bias, as we are dividing both two-factor interactions by 3.
However, even if we had knowledge that the two-factor interactions were of the same sign, we would not know the direction of the bias in this case without knowing the relative magnitudes of the two-factor interactions.

When estimating the main effect for factor 2, we would naturally approximate a different design.
Using the same logic as before, we would remove treatment combination $\bm{z}_5$.

As an alternative design, we may alias our main effects with the highest-order interaction possible, allowing for a different design for each main effect estimator.
So, when estimating the main effect of factor 1, we would use a fractional factorial design that aliases factor 1 with the three-factor interaction.
This leads to a design using treatment combinations $\bm{z}_1$, $\bm{z}_4$, $\bm{z}_5$, and $\bm{z}_8$ for which we can build the following estimand:
\begin{align*}
\dot{\tau}(1)&=\frac{1}{2}\left[\bar{Y}(\bm{z}_5) + \bar{Y}(\bm{z}_8)\right]-\frac{1}{2}\left[\bar{Y}(\bm{z}_1) + \bar{Y}(\bm{z}_4)\right]\\
&=\frac{1}{2}\frac{1}{2}\left(\bm{h}_8 +\bm{h}_5 - \bm{h}_4 - \bm{h}_1\right)\bm{\tau}\\
&=\tau(1)+\tau(123).
\end{align*}
Based on the hierarchy principle, an estimator with this aliasing structure should be a superior estimator to the original fractional factorial estimator mentioned because the three-factor interaction is more likely to be negligible than the two-factor interactions.
In particular, using the previous numbers from our empirical example, the bias for the main effect of pesticide A would be $1/3$.
In general, based on the hierarchy principle, we want to choose a design that aliases the effect we are looking at with the highest-order interaction possible (which may be limited by the missing data structure).
For example, if we are only missing data for one treatment combination, when estimating main effects we can always choose the $2^{K-1}$ fractional factorial design that aliases the main effect with the (negative of) the highest-order interaction.
This is not a design we would want to use to collect data, because it would alias our main effects with each other.
However, differing from the fractional factorial setting, we have more data and would use a different design for estimating each factorial effect.

Denote by $\dot{\bm{g}}_k$ the analog of $\bm{g}_k$ with zeros where outcome data are missing or excluded in a given estimator for factor $k$.
If there is a single treatment combination with no outcome measured, we can choose the aliasing such that factor $k$ is aliased with the $K$-factor interaction.
When more rows are missing, the pattern of missingness will dictate what aliasing structure is possible.
For example, if we are missing two treatment combinations but they are missing from the same design that aliases factor $k$ with the positive of the $K$-factor interaction, we can still recreate the design that aliases factor $k$ with the negative of the $K$-factor interaction.
But if we are missing a row from each of these designs, neither option is possible and we must choose a different aliasing structure.
If this method is continued for each factor, one ends up with a set of $\dot{\bm{g}}_k$, each with zeros for different treatment combinations.

Section 4.5 of \citet{WuHamada} gives a general strategy to design experiments while attempting to reduce aliasing for certain main effects.
There are other designs we can construct, not considered here, such as nonregular designs that have partial aliasing \citep[see][Chapter 7 for more details on nonregular designs]{WuHamada}.

\subsubsection{Statistical inference}
Once we pick which subsets of treatments we will use (based on the aliasing structure and practical constraints), we assume units are assigned using complete randomization to the $\tilde{J}$ treatments in the incomplete design.
Let $\bm{W}$ be the $n \times 2^{K}$ matrix of $W_i(\bm{z}_j)$ for all $i$ and $\bm{z}_j$, and $\bm{w}$ is some $n \times 2^{K}$ matrix with entries in the set $\{0,1\}$.
If $\tilde{\mathcal{J}}$ is the set of indices for the $\tilde{J}$ treatments included in the design, then
\[P(\bm{W} = \bm{w})= 
\prod_{j \in \tilde{\mathcal{J}}}n_j!/n! \]
if $\sum_{j \in \tilde{\mathcal{J}}}w_i(\bm{z}_j) = 1$  for all  $i$ and $\sum_{i=1}^nw_i(\bm{z}_j) = n_j$  for all  $\bm{z}_j$, where $n_j = 0$ for any treatments $\bm{z}_j$ not included in this design, and $P(\bm{W} = \bm{w}) = 0$ otherwise.
Assignment is similar to the full factorial or fractional factorial, just for a different subset of treatments.
The major differences between a fractional factorial design and incomplete design are therefore not in the way that the treatments are randomized but in (i) that the treatment subset used is not (in general) a fractional subset of the original $2^K$ treatments and (ii) the way in which the analysis is proposed to proceed.
All of the following inference results are derived under the assumption of complete randomization.

More generally, denote our estimator for $k^{th}$ factor under one of these alternative incomplete factorial designs as $\hat{\dot{\tau}}(k)=\dot{\bm{g}}_k^T\bar{\bm{Y}}^{obs}$.
It is straightforward to extend the variance expression $\text{Var}\left(\hat{\dot{\tau}}(k)\right)$ and variance estimator $\widehat{\text{Var}}\left(\hat{\dot{\tau}}(k)\right)$, as well as the covariance of $\hat{\dot{\tau}}(k)$ and $\hat{\dot{\tau}}(k')$, from Section~\ref{subsec:frac_fac_stat_anal}; in fact, it is easy to similarly extend these results to any linear combinations of interest on the $2^K$ treatment combinations.
However, terms for some treatment combinations will be set to zero in these expressions because they are present in one estimator but not the other.
See Appendix~\ref{append:partial_des} for the specification and derivation.

If we run a regression with all interactions on a dataset with missing treatment combinations, it is ambiguous what design the resulting estimators correspond to and, as discussed above, multiple designs may be plausible to estimate effects well.
If we specify a regression of the outcome on the fully interacted treatments, but some treatment combinations are not present in the data, the regression will not be able to estimate all interactions.
Not including a set of interactions implies that the linear model assumes that those interaction effects are zero, and we can assess how reasonable this assumption is.
However, the specific aliasing structure between the effects included in the model and those dropped by the regression is not obvious from the usual computer output alone, though can be discerned from the design matrix.
\citet{byar1993incomplete} and \citet{byar1995identifying} give more discussion of these types of estimators and Appendix~\ref{reg_miss} has further discussion.

\subsection{Fractional factorial designs vs incomplete factorial designs}

Sometimes an experimenter may have a choice between running a fractional factorial design or an incomplete factorial design.
However, in many cases this decision will be dictated in part by the ability to assign at least 2 units to each treatment in the design.
The fractional factorial design strictly uses $2^{K-p}$ of the original treatments, whereas an incomplete design may use more treatment groups.

When either design may be run, we expect there to be an advantage in terms of bias by using an incomplete factorial design.
In the incomplete factorial example given in Table~\ref{tab:partial_fac_example}, we were able to show that we can alias the main effect of factor 1 with the three factor interaction.
In our numerical example with the three pesticides, this implied a bias for our main effect estimator of pesticide A of $1/3$.
On the other hand, the fractional factorial design of Table~\ref{tab:frac_fac_design_example} resulted in factor 1 being aliased by the interaction between factors 2 and 3.
In our example, this meant the estimator for main effect of pesticide A was biased by 2.
We expect this larger bias by aliasing with two-factor interactions over three-factor interactions in general due to the hierarchy principle. 

In terms of variance, consider the case where we have the same total sample size for each design, a balanced design, and additive effects.
The fractional factorial design uses all $2^{K-p}$ treatment groups in its estimators of effects, and thus all $n$ units.
The incomplete design will only use a subset of the treatment combinations for estimating each effect, and therefore less than the total $n$ units.
It should be clear that we expect lower variance under the fractional design, in which more units are used in estimation.
Therefore, there is a bias-variance trade-off comparing the fractional factorial to the incomplete factorial design.

If instead we are comparing using different estimators (that correspond to different hypothetical design) after an incomplete design, such as in Table~\ref{tab:partial_fac_example}, was run, it should be clear that if the design is approximately balanced, we should expect similar variances for estimators that use the same number of treatments.

\section{Discussion: Embedding observational studies in fractional factorial designs}\label{sec:obs_stud}
\subsection{General issues}

To address causality in a non-randomized study, it has been argued that one needs to conceptualize a hypothetical randomized experiment that could correspond to the observational data \citep[][]{Bind17, rosenbaum2002observational, rubin2008objective, stuart2010matching}.
The hypothetical randomization is plausible if 
the treatment groups are ``similar'' with respect to confounding variables \citep{rubin2007design, rubin2008objective}.
With multiple treatments of interest, we would aim to recreate a hypothetical factorial randomized experiment.
However, in observational studies, there may be no units that received certain treatment combinations.
Therefore, we propose to recreate a hypothetical fractional factorial or incomplete factorial experiment, which further motivates the usefulness of these designs beyond randomized experiments.

For the fractional factorial design, we must decide which fraction to use.
This decision should be based on criteria such as the maximum resolution criteria.
To reduce the amount of aliasing a small $p$, i.e., a small fraction of total design being removed, is desirable.
In practice, we may not be able to control the aliasing structure. 
We must choose a $2^{K-p}$ design such that the treatment combination(s) with no observations is not used. 
However, this strategy usually results in the removal of units assigned to treatment combinations that were present in the observational data set but not used in the design.
If only one treatment combination is not present in the data set, we could use a $2^{K-1}$ design, but in that case we are not using $2^{K-1}-1$ treatment combinations for which we have data.

An alternative strategy to using a subset of the data would be to use an unsaturated regression model.
\citet{zhao2021regression} discusses this model from the potential outcome perspective, though not in the case where there are entire treatment groups missing from the data or containing only a single unit.
Those authors find that the unsaturated model will result in bias, in general settings, even with all treatment groups present.

After a design is chosen, a strategy to balance covariates should be used to ensure that the units across treatment groups are similar.
We assume strong ignorability; i.e., conditional on measured covariates, the assignment mechanism is individualistic, probabilistic, and unconfounded \citep{rosenbaum1983central}.
Then we can obtain unbiased causal estimates by analyzing the data as if it arose from a hypothetical randomized experiment after balancing.

Importantly, these assumptions apply to all treatment combinations in the final experimental design used.
We must also assume that the treatment assignment is, at least hypothetically, manipulable such that all potential outcomes are well-defined.
This ensures that our estimands of interest are defined.
See Supplementary Material~\ref{append:obs_stud} for more discussion.
A similar argument must hold for the unconfoundedness assumption.

\subsection{Covariate balance}\label{sec:cov_bal}

An important stage when estimating the causal effect of non-randomized treatments is the design phase \citep{rubin2007design,rubin2008objective}.
At this stage, we attempt to obtain a subset of units for which we can assume unconfoundedness of the treatment assignment.
That is, units for which $P(W_i|Y_i(z),\bm{X}_i)=P(W_i|\bm{X}_i)$, where $\bm{X}_i$ is an $m$-dimensional vector of covariates \citep{Imbens15}.
Matching strategies are often used to ensure no evidence of covariate imbalance between treatment groups, as reviewed by \citet{stuart2010matching}.

With multiple treatments, matching can be difficult and methods for obtaining covariate balance specifically for factorial type designs require further exploration.
There are extensions of propensity score balancing to multiple treatments, notably the generalized propensity score (GPS) \citep{hirano2004propensity} and the covariate balancing propensity score (CBPS) \citep{imai2014covariate}.
\citet{Lopez17} review techniques, including matching, for observational studies with multiple treatments and \citet{bennett2018building} uses template matching, which matches units to a ``template'' population of units, for multiple treatments.
Although neither of those works explore factorial (full or fractional) designs, it may be straightforward to extend their methods to this setting.
\citet{nilsson2013causal} discusses matching in the $2^2$ design.
In our data illustration in Section~\ref{sec:app}, we employ sequential trimming and checks on covariate balance.
Tests for covariate balance are reviewed in Supplementary Material~\ref{sec:cov_bal_test}.

One issue likely to occur with matching is the difficulty in achieving good balance across many groups.
This could result in dropping a large number of units, leading to lower power to detect effects.
Weighting techniques can avoid the issue of completely dropping units, but the weights may be unstable if there is little overlap between all treatment groups.

Due to challenges in obtaining full balance, a first step might be to only obtain covariate balance between two treatment groups, a task commonly done in the causal inference literature, and compare outcomes for these groups.
To start, we could estimate the difference in outcomes between the units that were assigned level $+1$ for all factors and the units that were assigned $-1$ for all factors.
Under certain assumptions, testing the difference in these groups can act as a global test for whether any effects of interest are significant.
We discuss the test and assumptions that are needed for this simple comparison to be meaningful in Supplementary Material~\ref{subsec:test_sig_eff}.

\subsection{Comparing designs}\label{sec:design_obs_comp}
There may be complications in the observational setting that make attempting to recreate one design easier or more desirable than another.
If we use a different design for each estimator, as in Section~\ref{sec:part_fac}, we are able to use more of the data than a fractional factorial design, but this can also incur a cost.
If we have no observations for only one treatment combination, we would use all $2^K-1$ treatment combinations if we did an analysis for all effects and used a different design for each.
This approach would require us to either first obtain balance among all $2^K-1$ treatment groups or to obtain balance among the treatment groups within each design separately.
The former option may be difficult; as the number of treatment combinations grows, obtaining covariate balance across all treatment groups becomes increasingly difficult and may result in smaller and smaller sample sizes, especially if trimming is used.
The latter option will make joint inferences more challenging because different units would be used in each analysis.
Therefore, although these incomplete factorial designs may improve the bias of our estimators, the fractional factorial design in which we are using the same $2^{K-p}$ rows may be more attractive in terms of obtaining covariate balance.

Further, estimators for different designs result in different types of bias due to aliasing of effects.
In order to believe that the biases of main effect estimators are negligible, we must believe that the interactions aliased with that main effect are negligible.
Researchers need to carefully unpack the aliasing structure for the design they choose in the observational setting in order to understand the assumptions necessary for unbiased estimation.
This requires even more care and thought than a standard observational study.

\subsection{Data illustration}\label{sec:app}
\subsubsection{Data description}

We now give an illustration of the implementation of our methods using data on pesticide exposure and body mass index (BMI), the ratio between weight and height-squared.
Previous studies have found an association between pesticide exposures and body mass index (BMI) \citep{Buser14, Ranjbar15}.
We use the 2003-\citeyear{nhanes} cycle of the National Health and Nutrition Examination Survey (NHANES) collected by the Centers for Disease Control and Prevention (CDC).
We access the data via the \texttt{R} \citep{R_cite} package \texttt{RNHANES} \citep{rnhanes}. 
We focus on four organochlorine pesticides, measured via a blood serum test and then dichotomized based on whether they were above (+1) or below (-1) the detection limit, as given in the NHANES dataset, as factors.
Organochlorine pesticides are persistent in the environment and adverse health effects have been reported by the CDC (\citeyear{nhanesorganexp}), making them an interesting group of pesticides to study.

We focus on four pesticides (beta-Hexachlorocyclohexane (beta-Hex), heptachlor epoxide (Hept Epox), mirex, and p.p'-DDT), chosen primarily based on data availability and exposure rates.
We removed 271 units with missing values of pesticide and BMI, noting that because this is an illustration and not intended to draw causal conclusions we simply drop those units.
We decided to study a non-farmer population as farmers are more likely to be exposed to pesticides than the general population and may also differ on other unobserved covariates that affect health outcomes.
This is our first step to achieving covariate balance, leaving a dataset with 1,259 observations (see Figure~\ref{fig:farmer_bal_bmi} in the Appendix).
Importantly, as shown in Table~\ref{tab:data_fullfrac}, we have one treatment group with only one unit.
Thus, using the full dataset can lead to unstable estimates and an inability to calculate Neyman variance.

\subsubsection{Design stage}\label{sec:data_ex_design}
To show the process of how estimands change as we adjust our design stage, we consider three different hypothetical ``experiment'' designs:
1) a 2$^4$ factorial design, without adjusting the sample to balance for covariates
2)  a fractional factorial 2$^{4-1}$ design, without adjusting the sample to balance for covariates 
3) a fractional factorial 2$^{4-1}$, with trimming to obtain covariate overlap and no evidence of covariate imbalance with respect to gender (recorded as male vs. female), smoking status (smoker vs. non-smoker), and age at the time of survey (in years).
Comparing designs (1) and (2) will show how estimates change between a full factorial and a fractional factorial design.
Comparing designs (2) and (3) will show how estimates change when a design phase with the aim of obtaining covariate balance is implemented.
We use simple trimming as the focus of this paper is not to compare different methods of balancing for covariates, though exploration of different balancing methods in this setting would be worthwhile.
We consider gender, age, and smoking status as our first tier of most important covariates.
For smoking status, we use the question ``Have you smoked at least 100 cigarettes in your life time?'' and categorize ``Yes'' and ``Don't know'' as smokers.
Only one observation with value ``Don't know'' was recorded.

\subsubsection{Statistical analysis}

We analyzed the three datasets described in the design stage using:
1) a multiple linear regression that regresses BMI on treatment factors;
2) a multiple linear regression that regresses BMI on treatment factors, as well as the following covariates, as factors: race and ethnicity, income, gender, and smoking status;
3) Fisher-randomization tests of the sharp null hypothesis of no treatment effects.
For the first analysis, recall that regression estimates when including all factors and interactions, but not covariates, correspond to the Neymanian estimates divided by two. 
The second analysis is motivated by Figures~\ref{fig:ethnicity_after_match} and \ref{fig:income_after_match} in Appendix~\ref{append:frac_fac_mat_des}, which show that further adjustment for ethnicity and income are needed, even after trimming.
The second analysis also adjusts for our second tier of covariates: race and ethnicity and income.
Income is defined as annual household income.
From now on, we will refer to the ``race and ethnicity'' covariate as simply ethnicity, as a short hand and to make clear that this is one categorical variable in the NHANES dataset.
In the second analysis for each design, units with missing covariate values were removed.
Due to the right-skewed nature of the weight variable, BMI exhibits some degree of right-skewness.
Therefore, we use log-transformed BMI as the outcome in all of these analyses.
Additional data descriptions and the full analyses are available in Appendix~\ref{append:data_ill}.

\subsubsection{Resulting designs}
Table~\ref{tab:data_fullfrac} provides the counts of observations for each treatment ($\bm{z}_j$).
Factor combination 10 ($\bm{z}_{10}$) has only one observation.
Relying on only one observation for a treatment will lead to unstable estimates and we will not be able to use Neymanian variance estimates.
Hence, the fractional factorial design aims to avoid this issue by embedding the observational study in a fractional factorial hypothetical experiment.

\def\spacingset#1{\renewcommand{\baselinestretch}
{#1}\small\normalsize} \spacingset{1}
\begin{table}[ht]
\centering
\begin{tabular}{rcccccc}
\hline
& \multicolumn{4}{c}{Factor Levels}&\multicolumn{2}{c}{Number of Obs.}\\
  \hline
  & \textbf{beta-Hex} & \textbf{Hept Epox} & \textbf{Mirex} & \textbf{p,p'-DDT} & \textbf{Original} &  \textbf{Trimmed}\\ 
  \hline
$\bm{z}_1$ & \phantom{-}+1 & \phantom{-}+1 & \phantom{-}+1 & \phantom{-}+1  & 426 & -\\ 
 \color{red} $\bm{z}_2$ & \phantom{+}-1 & \phantom{-}+1 & \phantom{-}+1 & \phantom{-}+1 &   12 & 6\\ 
 \color{red} $\bm{z}_3$ & \phantom{-}+1 & \phantom{+}-1 & \phantom{-}+1 & \phantom{-}+1 &   70 & 22\\ 
  $\bm{z}_4$ & \phantom{+}-1 & \phantom{+}-1 & \phantom{-}+1 & \phantom{-}+1 &  51 & -\\ 
  \color{red}$\bm{z}_5$ & \phantom{-}+1 & \phantom{-}+1 & \phantom{+}-1 & \phantom{-}+1 & 291  & 102\\ 
 $\bm{z}_6$& \phantom{+}-1 & \phantom{-}+1 & \phantom{+}-1 & \phantom{-}+1 &  25  & -\\ 
 $\bm{z}_7$ & \phantom{-}+1 & \phantom{+}-1 & \phantom{+}-1 & \phantom{-}+1& 94  & -\\ 
\color{red}  $\bm{z}_8$ & \phantom{+}-1 & \phantom{+}-1 & \phantom{+}-1 & \phantom{-}+1 & 54  & 10 \\ 
 \color{red} $\bm{z}_9$ & \phantom{-}+1 & \phantom{-}+1 & \phantom{-}+1 & \phantom{+}-1 &  21  & 8  \\ 
  $\bm{z}_{10}$ & \phantom{+}-1 & \phantom{-}+1 & \phantom{-}+1 & \phantom{+}-1 & 1 & - \\ 
  $\bm{z}_{11}$ & \phantom{-}+1 & \phantom{+}-1 & \phantom{-}+1 & \phantom{+}-1 &  19 & - \\ 
\color{red}  $\bm{z}_{12}$ & \phantom{+}-1 & \phantom{+}-1 & \phantom{-}+1 & \phantom{+}-1 & 19 & 10 \\ 
  $\bm{z}_{13}$ & \phantom{-}+1 & \phantom{-}+1 & \phantom{+}-1 & \phantom{+}-1 &     42  & - \\ 
\color{red}  $\bm{z}_{14}$ & \phantom{+}-1 & \phantom{-}+1 & \phantom{+}-1 & \phantom{+}-1 &   19 & 3  \\ 
\color{red}  $\bm{z}_{15}$ & \phantom{-}+1 & \phantom{+}-1 & \phantom{+}-1& \phantom{+}-1 &   37 & 8 \\ 
  $\bm{z}_{16}$& \phantom{+}-1 & \phantom{+}-1 & \phantom{+}-1 & \phantom{+}-1& 78 & - \\ 
   \hline
   &$\bm{g}_1$&$\bm{g}_2$&$\bm{g}_3$&$\bm{g}_4$&1259 & 169\\
   \hline
\end{tabular}
\caption{Design matrix and counts.
+1 refers to exposure to the pesticide, -1 refers to no detectable exposure to the pesticide.
Red treatments are used when recreating a fractional factorial design.
``Original'' column on the right gives the counts for each treatment for the full data set.
``Trimmed'' column on the right gives counts for those treatments used in the fractional factorial design, after trimming.
}
\label{tab:data_fullfrac}
\end{table}
\def\spacingset#1{\renewcommand{\baselinestretch}
{#1}\small\normalsize} \spacingset{1.5}

Recalling that $I$ corresponds to the intercept, in the $2^{4-1}$ fractional factorial design, instead of using $I = 1234$, we chose $I=-1234$ to exclude row 10 in Table~\ref{tab:data_fullfrac} that has only a single observation.
The dataset in this hypothetical experiment consists of 523 observations.
In this design, aliasing is as follows: $I = -1234$, $4=-123$, $3=-124$, $2=-134$, $1=-234$, $12=-34$, $13=-24$, $14=-23$.
The main effects are aliased with the negative of the three-factor interactions and the two-factor interactions are aliased with each other with reversed signs.
In order to identify main effects, we will assume that the three-factor interactions are negligible.
In practice, researchers should assess whether this aliasing assumption is realistic.

For the third design we attempted to get balance on the covariates of gender, smoking status, and age.
Figure~\ref{fig:cov_bal_bmi} shows covariate imbalance with respect to gender, smoking, and age in the fractional factorial design before trimming.
Recall that these form our first tier of covariates and ethnicity and income constitutes the second tier, which is adjusted for via linear regression.
In practice, balancing a large set of covariates is not always possible and so choosing the most important (first tier) covariates should be done using subject matter knowledge.
We used a rejection approach that sequentially pruned observations from the fractional factorial dataset until we found no evidence of imbalance across exposure groups with respect to our first tier of covariates, based on a MANOVA using the Wilks' statistic \citep[][see also Supplementary Material~\ref{sec:cov_bal_test}]{wilks1932certain}.
Figure~\ref{fig:cov_bal_bmi_after_trim} shows the covariate distribution for our first tier after trimming.
Gender and smoking status are still imbalanced even after trimming, and hence were adjusted for in the linear regression that includes covariates (this is true for all ``designs'').
The first dataset that resulted in no evidence of covariate imbalance consisted of 169 observations, and the new treatment counts are presented in Table~\ref{tab:data_fullfrac}.
The number of units has been drastically reduced in our attempt to achieve covariate balance, a major challenge in this setting.
In fact, one treatment group only has three observations.

\subsubsection{Results comparison across different conceptualized experiments and statistical approaches}

Figure~\ref{fig:compare_est} shows a comparison of the regression estimates of the main effects across designs and statistical analyses.
To compare the different methods we present univariate analyses.
That is, we utilize individual tests for each main effect rather than joint tests, for better illustration of the different methods.
In practice, adjustment for multiple comparisons should be considered.
All methods and designs estimated a positive ``effect'' of heptachlor epoxide and a negative ``effect'' of mirex on BMI.
Although the full factorial estimates generally agree with the estimates of the two fractional factorial designs, differences in estimates of beta-Hexachlorocyclohexane (Beta-Hex) and p,p'-DDT may be due to the aliasing of the three-factor interactions with the main effects. 
However, it is also plausible that we have reduced our data in the fractional design to a subset of individuals with different average main effects than in the full data set.

Figure~\ref{fig:compare_sig} shows a comparison of the significance of these estimates. The Fisher p-value is the p-value for the test of no effects of any pesticides, based on effect estimates for a given pesticide, which is suggested as a screening stage in \citet{espinosa2016bayesian}.
We obtained low p-values for the main effect of mirex on BMI across all methods and designs.
However, the p-values disagree for the other pesticides, especially the p-values testing the effects of p,p'-DDT and beta-Hexachlorocyclohexane.
The HC2/Neyman p-value is the significance based on the HC2 variance estimate (or Neyman variance estimate as we have shown this estimator to be equivalent in settings with no covariates) and the normal approximation.
This p-value was only calculated for the regression analysis without covariates and was unavailable for the full factorial model due to limited data.

\subsubsection{Discussion of data illustration}
We have performed a data illustration to show the benefits and challenges of using our method and working with observational data with multiple treatments in general.

First, it is important to note that simplifications were made in the statistical analyses to focus on illustrating how researchers can capitalize on using fractional factorial designs to estimate the main and interactive effects of multiple treatments in observational studies.
We did not adjust for all important hypothetical covariates, such as diet.
Another consideration is that we log-transformed BMI to address the fact that BMI is a ratio and so its distribution tends to have heavy tails.
However, heavy-tailed distributions could have been considered. 

There were also some major challenges in working with this data and our method related to sample size. Trimming helps mitigate bias but greatly reduced our sample size, potentially leading to decreased power and precision.
There was also a great reduction in sample size by using the fractional factorial design which entailed dropping treatment combinations.
We could have used an incomplete factorial type design, but this likely would have made the covariate balancing even more challenging, as discussed previously.

For p,p'-DDT, the full factorial design resulted in a lower p-value than the other designs, which could be an issue of aliased interactions watering down the effect in the fractional design, as a result of the bias noted in Result~\ref{result:frac_bias}.
It could also have occurred because the populations are different in these two designs.
For beta-Hexachlorocyclohexane, the covariate balance adjusted fractional factorial design has a lower p-value compared to the other designs.
We are more inclined to trust the balanced design as this should have reduced bias.
This result may indicate that there was some confounding that made the effect appear less significant before trimming.
In fact, the stark contrast between the unbalanced and balanced fractional design suggests that confounding may be to blame.
Alternatively, by trimming we may have reduced our sample to a subpopulation where beta-Hexachlorocyclohexane has a larger effect on BMI than the rest of the population.

 It is important to acknowledge that using the small subset of the NHANES dataset, we do not intend to provide policy recommendations on pesticide use.
In the general population, organochlorine pesticide exposure primarily occurs through diet (excluding those with farm-related jobs), particularly eating foods such as dairy products and fatty fish \citep{nhanesorganexp}.
Without further adjustments for diet, we are not be able to disentangle the causal effect of diet and pesticides.
For instance, in our study individuals are likely to have been exposed to mirex largely through fish consumption (Agency for Toxic Substances and Disease Registry [ATSDR], \citeyear{mirexfish}).
Further studies could investigate BMI differences in a group of fish consumers with high levels of mirex and a ``similar'' group of fish consumers with low levels of mirex, where similar is with respect of important confounding variables. 
Indeed, it could be that eating fish causes individuals to have both high levels of mirex and also lower BMI.

\section{Epilogue}\label{sec:conc}

In this paper, we have laid out how to design and analyze fractional factorial and incomplete factorial designs using the potential outcomes framework.
These designs are particularly useful when running a full factorial design is infeasible or impractical, but the factorial estimands are still of interest.
Our work includes extensions of some of the known factorial design results for variance and regression to the fractional factorial and incomplete design setting.
In doing so, we have expanded the tool kit available to researchers wishing to perform causal inference.

Further, we have proposed to embed observational studies with multiple treatments in fractional factorial hypothetical experiments.
This type of design is useful in settings with many treatments, especially when some treatment combinations have few or no observations and the aliasing assumptions are plausible.
Once we recreate a factorial or fractional factorial experiment in the design phase, we can use standard methods, extended as in Sections~\ref{sec:full_fac} and \ref{sec:frac_fac}, to estimate causal effects of interest. 
We first reviewed the basic setup for factorial and fractional factorial designs.
We have also discussed covariate balance complications that may arise when dealing with multiple non-randomized treatments in practice
and illustrated these methods on a data set with pesticide exposure and BMI.

We have given a short overview of factorial, fractional factorial, and incomplete factorial designs, in the potential outcomes framework.
However, there are many related designs that we did not cover \citep[see][]{WuHamada}, such as nonregular design types (e.g., Plackett-Burman designs).
Practitioners may also use variable selection and the principle of effect heredity to select their model for estimating factorial effects.

We see many avenues of future exploration connected to our approach.
For instance, coupling fractional factorial designs with a Bayesian framework would provide more statistical tools and would potentially offer different methodology for dealing with missing data.
Additionally, development of observational tools, especially covariate balancing techniques for factorial designs with many treatment combinations should be an area of future exploration.
A particular challenge is that as we increase the number of treatment combinations, matching becomes more and more difficult due to the increased dimensionality and weighting methods may produce unstable estimators.
One direction could also be to choose the design of the observational study based upon the ability to balance different treatment groups.
Random allocation designs \citep{dempster1960random, dempster1961random}, in which randomness of the design is incorporated, could also be utilized in this framework.
Finally, we could explore other causal estimands that may be of interest in observational studies with multiple treatments, such as those in \citet{egami2017causal} and \citet{de2019improving}.

\def\spacingset#1{\renewcommand{\baselinestretch}
{#1}\small\normalsize} \spacingset{1}
\bibliographystyle{apalike}
\begingroup
    \setlength{\bibsep}{4pt}
\bibliography{fracfacref}
\endgroup

\begin{appendix}
\include{figures}
\end{appendix}

\begin{appendix}
\include{fracfacappend}

\end{appendix}

\end{document}

%% file: figures.tex
\section*{Figures}

\begin{figure}[h!]
\centering
\includegraphics[scale=0.58]{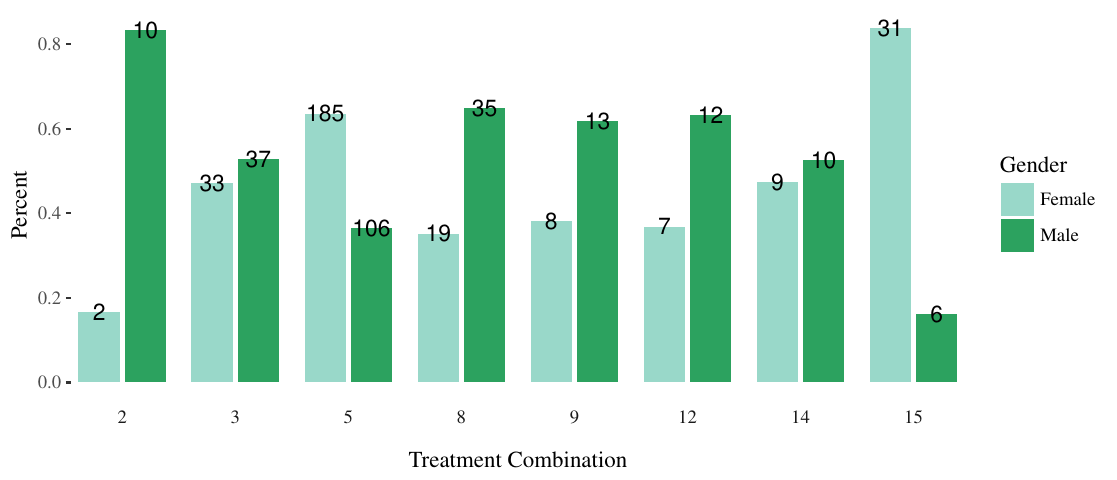}
\includegraphics[scale=0.58]{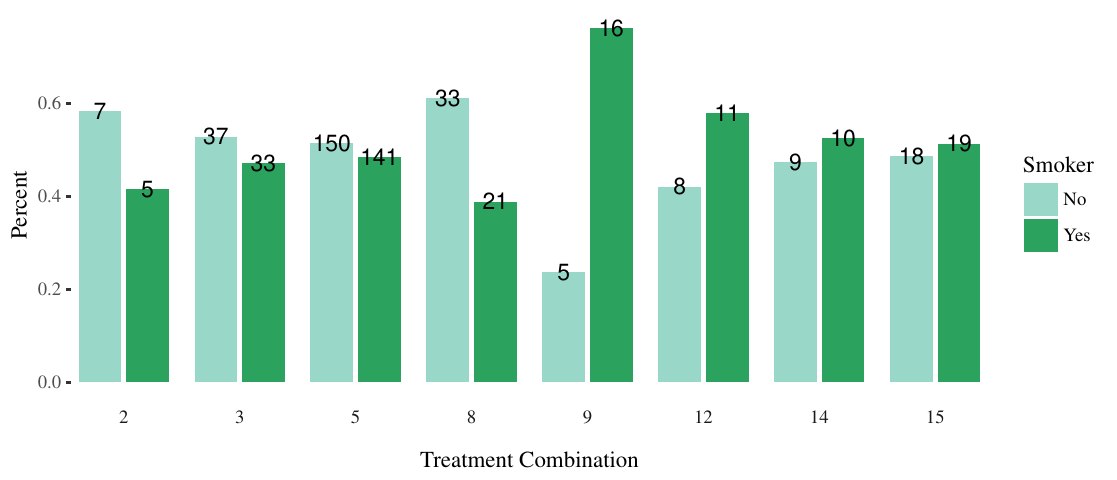}
\includegraphics[scale=0.58]{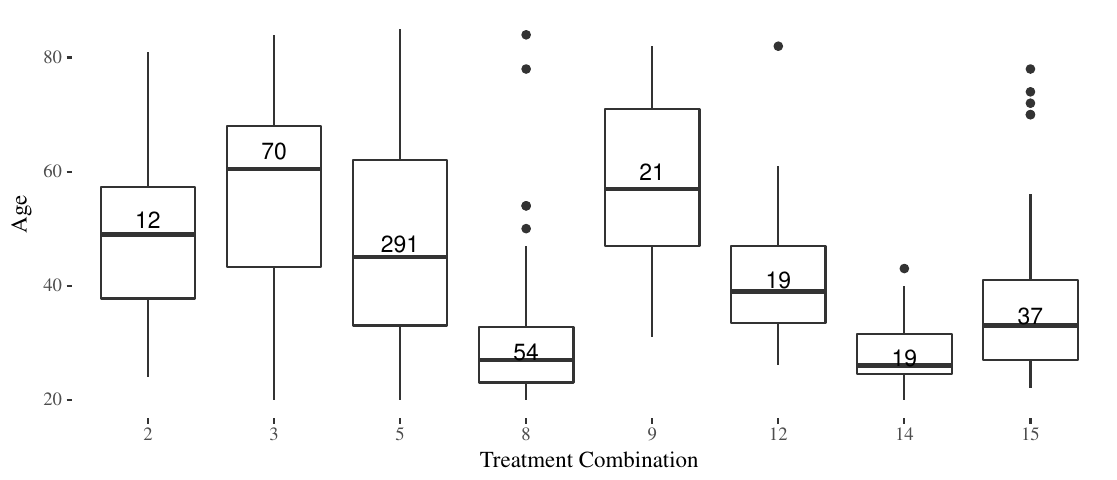}
\caption{Comparing covariates across treatment combinations in the $2^{4-1}$ fractional factorial design.
Text labels give number of observations per group.
For bar plots, the y-axis gives the percent within each treatment combination for each category of the covariate.
For age, individuals with ``$>=$ 85 years of age'' were set to 85 on the graph. Note that all individuals older than 85 were dropped in the covariate balance stage.}
\label{fig:cov_bal_bmi}
\end{figure}

\begin{figure}[h!]
\centering
\includegraphics[scale=0.58]{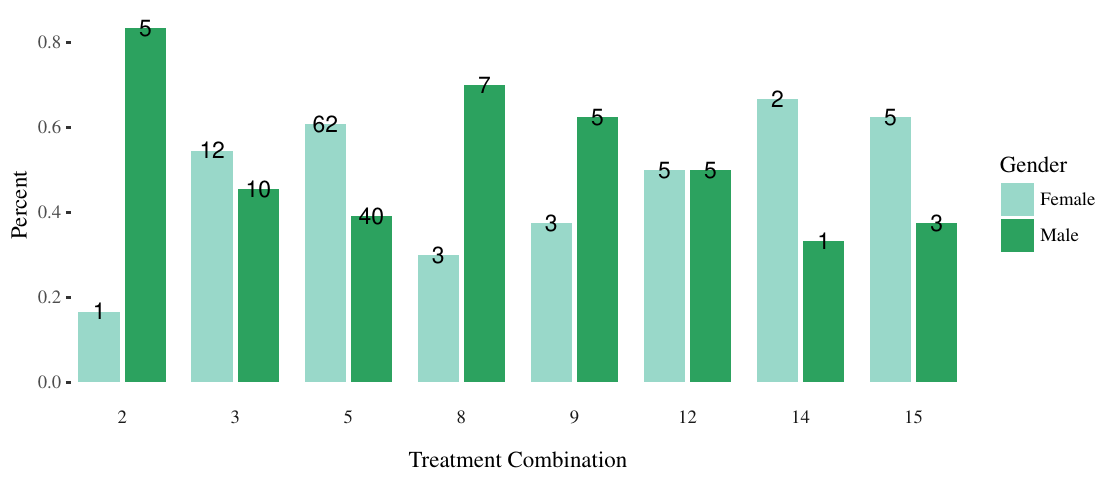}
\includegraphics[scale=0.58]{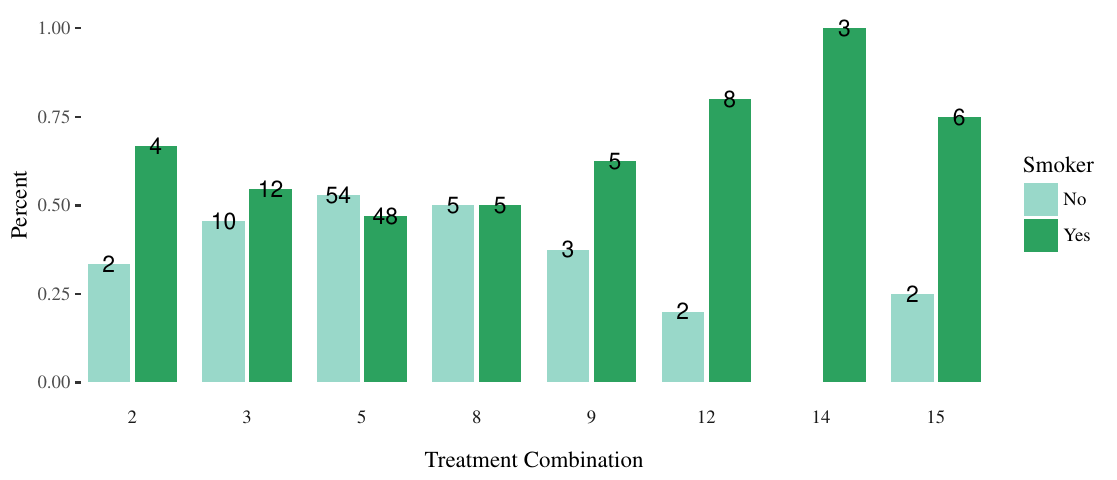}
\includegraphics[scale=0.58]{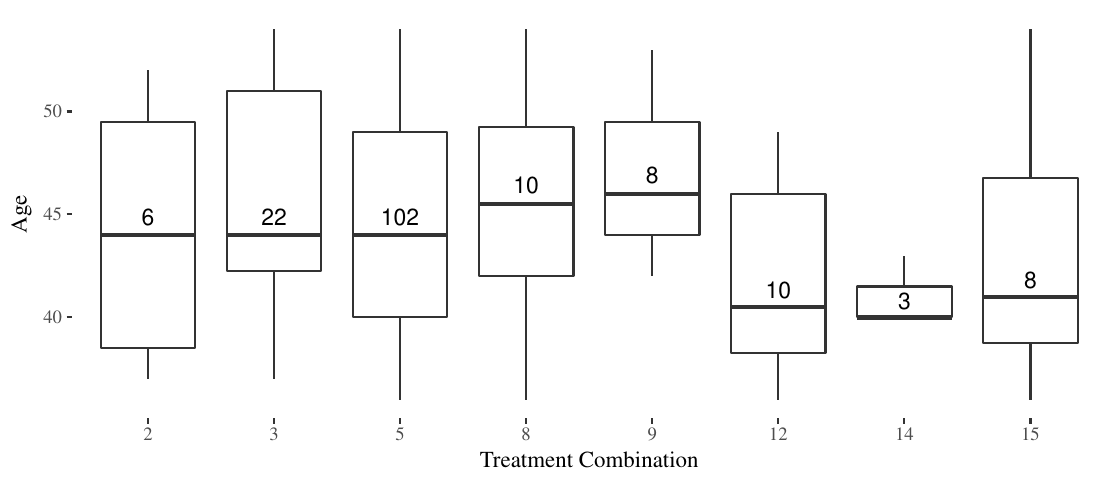}
\caption{Comparing covariates across treatment combinations in the $2^{4-1}$ fractional factorial design after trimming.
Text labels give number of observations per group.
For bar plots, the y-axis gives the percent within each treatment combination for each category of the covariate.
For age, individuals with ``$>=$ 85 years of age'' were set to 85 on the graph. Note that all individuals older than 85 were dropped in the covariate balance stage.
}
\label{fig:cov_bal_bmi_after_trim}
\end{figure}

\begin{figure}
\centering
\includegraphics[scale=0.8]{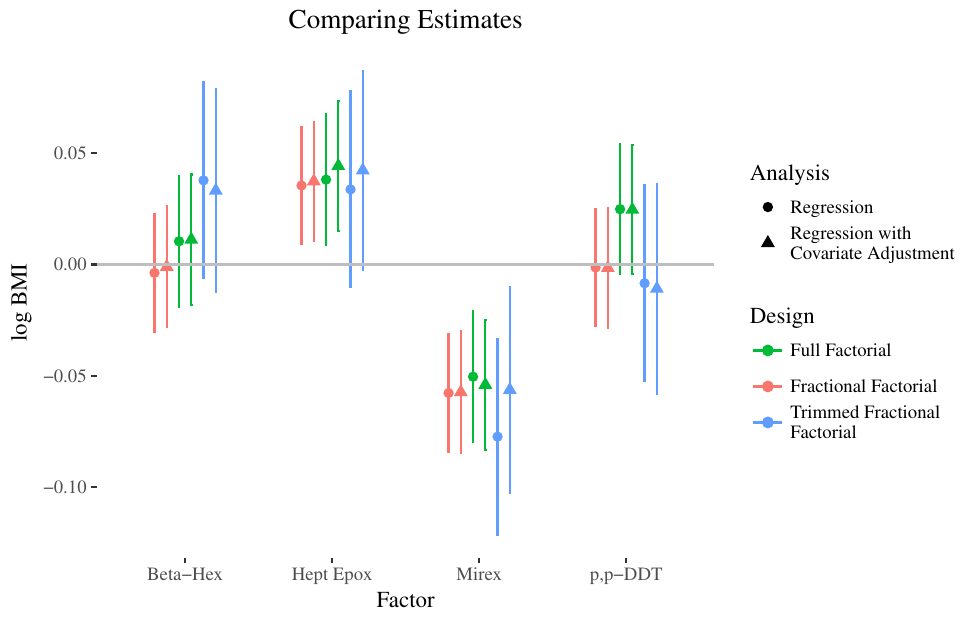}
\caption{Plots of estimates of factorial effects (associations), on the log BMI scale.
Bars indicate two standard errors (using standard OLS standard estimates because the Neymanian variance estimates are not available for the full factorial design due to having only a single unit in one treatment group) above and below point estimate.}
\label{fig:compare_est}
\end{figure}

\begin{figure}
\centering
\includegraphics[scale=0.8]{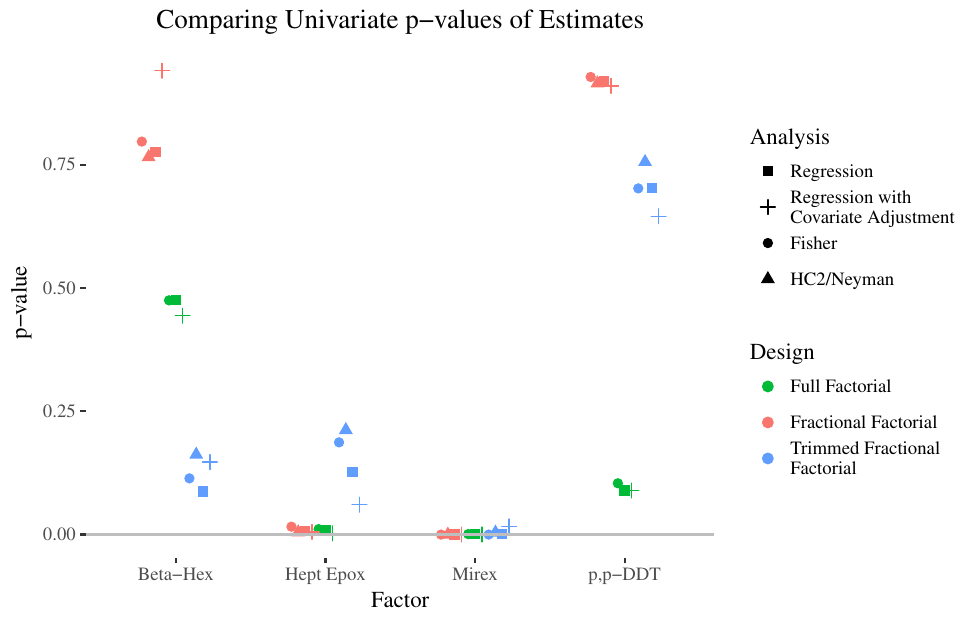}
\caption{Plots of p-values of factorial effect (association) estimates, which are compared in Figure~\ref{fig:compare_est}.}
\label{fig:compare_sig}
\end{figure}

%% file: fracfacappend.tex
\section{Statistical inference for full factorial}\label{sec:fac_stat_inf}
In this section we provide some further review of full factorial designs under Neyman-style randomization-based inference.
A conservative Neyman-style estimator for the variance was proposed by \cite{Dasgupta2015} and extended to the unbalanced case by \cite{lu2016randomization}:
\begin{align}
\widehat{\text{Var}}\left(\widehat{\tau}(k)\right) = \frac{1}{2^{2(K-1)}}\sum_{j=1}^J\frac{1}{n_j}s^2(\bm{z}_j),
\end{align}
where \[s^2(\bm{z}_j)=\frac{1}{n_j-1}\sum_{i:W_i(\bm{z}_j)=1}\left(Y_i(\bm{z}_j) - \bar{Y}^{obs}(\bm{z}_j)\right)^2\] is the estimated sampling variance of potential outcomes under treatment combination $\bm{z}_j$.
\cite{Dasgupta2015} discussed other variance estimators and their performance under different assumptions (e.g. strict additivity, compound symmetry).
In that paper, the authors also provided a Neyman-style estimator for the covariance by again substituting $s^2(\bm{z}_j)$ for $S^2(\bm{z}_j)$. 
Unfortunately, this estimator is not guaranteed to be conservative because $S^2_{k,k'}$ may be positive or negative.
The same authors provided a Neyman confidence region for $\bm{\tau}$, the vector of all $J-1$ factorial effects. First, they define, in Equation 26,
\[T_n = \hat{\bm{\tau}}^T\hat{\Sigma}_{\hat{\bm{\tau}}}^{-1}\hat{\bm{\tau}},\]
where $\hat{\bm{\tau}}$ is the vector of Neyman estimators of $\bm{\tau}$ that we defined earlier and $\hat{\Sigma}_{\hat{\bm{\tau}}}$ is the estimator of the covariance matrix of $\hat{\bm{\tau}}$, $\Sigma_{\hat{\bm{\tau}}}$.
Then the $100(1-\alpha)\%$ confidence region for $\bm{\tau}$ (Equation 27 of \citet{Dasgupta2015}), is
\[\{\hat{\bm{\tau}}:p_{\alpha/2} \leq T_n \leq p_{1-\alpha/2}\},\]
where $0<\alpha<1$ and $p_{\alpha}$ is the is the $\alpha$ quantile of the asymptotic distribution of $T_n$.
In that paper, a Neyman-style confidence interval for individual effects was also provided as
\[\hat{\tau}(k) \pm z_{\alpha/2}\sqrt{\widehat{\text{Var}}\left(\widehat{\tau}(k)\right)},\]
where the authors rely on a normal approximation for the distribution of $\hat{\tau}(k)$, typically assumed to hold asymptotically.
See Proposition 3 and Example 11 in \citet{li2018rerandomization} for conditions under which the asymptotic normality holds in this setting.

Note that a model that linearly regresses an outcome against the factors coded using contrast values $-1$ and $+1$ and all interactions between factors (but no other covariates) will result in the same point estimates for the factorial effects as presented here, divided by 2 \citep{Dasgupta2015, lu2016randomization}.
For a balanced design or when treatment effects are assumed to be additive, so that the variances $S^2(\bm{z}_j)$ are all the same, the standard linear regression variance estimate, relying on homoskedasticity, will be the same as the Neymanian variance estimate \citep{Dasgupta2015,Imbens15}.
However, this is not true for an unbalanced design.
\cite{samii2012equivalencies} showed that the HC2 heteroskedasticity robust variance estimator \citep[see][for more details]{mackinnon1985some} is the same as the Neymanian variance estimator presented here for a single treatment experiment.
\cite{lu2016randomization} extended this finding to factorial designs, showing that this estimator is also the same as the Neymanian variance estimator in that case.
Note that because the regression estimator is different by a factor of 2, this regression variance is different by a factor of 4.

Fisherian and Bayesian types of analyses are also possible.
See \cite{Dasgupta2015} for additional discussion.
In particular, they explore creating ``Fisherian Fiducial'' intervals \citep{fisher1930inverse, wang2000fiducial} for factorial effects.
The basic idea is to invert our estimands so that we write our potential outcomes in terms of the estimands.
Then, under a Fisher sharp null hypothesis, we can impute all missing potential outcomes and generate the randomization distribution.

\section{Tests for covariate balance}\label{sec:cov_bal_test}
Testing for covariate imbalance across multiple treatment groups can be done using multivariate analyses of variance (MANOVA), which  uses the covariance between variables to test for mean differences across treatment groups, as used in \citet{branson2016improving}.
Recall that the factorial design has $J$ treatment combinations.
Define the following H and E matrices \citep{Coombs96}:

\begin{center}

$$\bm{H}=\sum_{k=1}^{J} n_k (\overline{\bm{X}}_k-\overline{\bm{X}})(\overline{\bm{X}}_k-\overline{\bm{X}})^T $$

$$\bm{E}=\sum_{k=1}^{J} (n_k-1) S_k,\ \text{where} \ S_k=\frac{1}{n_k-1}\sum_{j=1}^{n_k} (\bm{X}_{kj}-\overline{\bm{X}}_k)(\bm{X}_{kj}-\overline{\bm{X}}_k)^T, $$

\end{center}
where $\overline{\bm{X}}_k$ is the $m$-dimensional vector of mean covariate values for treatment group $k$, $\overline{\bm{X}}$ is the average $m$-dimensional vector of mean covariate values for all units, that is $\overline{\bm{X}} = \sum_{k=1}^{J}\frac{n_k}{n}\overline{\bm{X}}_k$, and $\bm{X}_{kj}$ is the $m$-dimensional vector of covariates for the $j$th unit in treatment group $k$.
Denote by $\theta_k$ the ordered eigenvalues of $\bm{H}\bm{E}^{-1}$, where $k \in\{1,...,s\}$ and $s=\min(m,K-1)$.
Standard MANOVA statistics, which can be used to test covariate balance, are typically functions of the eigenvalues of $\bm{H}\bm{E}^{-1}$ \citep{Coombs96}.
We chose the Wilks' statistic \citep{wilks1932certain},

$$\text{Wilks}=\frac{|\bm{E}|}{|\bm{H}|+|\bm{E}|} =\prod_{k=1}^{K}\frac{1}{1+\theta_k},$$
where $\theta_k$ corresponds to the $k^{th}$ eigenvalue of $\bm{H}\bm{E}^{-1}$.

As discussed in \citet{imai2008misunderstandings}, a potential drawback of testing for evidence against covariates imbalance is that as we drop units we lose power to detect deviations from the null hypothesis of no difference in covariates between the treatment groups.
Another diagnostic for covariate balance is checking covariate overlap via plots and other visual summaries of the data.
So called ``Love plots'', which show standardized differences in covariate means between two treatment groups before and after adjustment \citep{ahmed2006heart}, are difficult to generalize directly because of the multitude of treatment groups and comparisons.
However, plots of standardized means or of distributions may be helpful to detect imbalance.

\section{Initial test for significance of effects}\label{subsec:test_sig_eff}

As discussed in the previous section, achieving covariate balance for more than two treatment groups can be a challenge.
Therefore, instead of attempting to achieve balance among all treatment groups, a simple first step might be to examine two carefully chosen treatment groups and attempt to balance these two groups only.
Obtaining balance between two treatment groups has been well-studied in causal inference; see \citet{stuart2010matching} for a review of common matching methods.
Once significant covariate imbalance can be ruled out, we can test whether the mean difference between these two groups is significantly different.
But what can we learn from this comparison about our factorial effects?

We have from Section~\ref{subsec:est_fac} that
\[\bar{Y}(\bm{z}_j) = \frac{1}{2}\bm{h}_j\bm{\tau}.\]
So when we subtract two observed means, assuming that the observed means are unbiased estimates of the true means (i.e. we have randomization or strong ignorability), we are estimating
\[\bar{Y}(\bm{z}_j) - \bar{Y}(\bm{z}_{j'})= \frac{1}{2}(\bm{h}_j-\bm{h}_{j'})\bm{\tau},\]
which is the sum of terms that are signed differently in $\bm{h}_j$ and $\bm{h}_{j'}$.

As an example for a $2^3$ design, we have the following matrix for $\bm{G}^T$:

\def\spacingset#1{\renewcommand{\baselinestretch}
{#1}\small\normalsize} \spacingset{1}
\begin{align*}
\bm{G}^T &=\begin{bmatrix}
\bm{h}_1\\
\bm{h}_2\\
\bm{h}_3\\
\bm{h}_4\\
\bm{h}_5\\
\bm{h}_6\\
\bm{h}_7\\
\bm{h}_8\\
\end{bmatrix}
=
\begin{pmatrix*}[r]
+1 & -1 & -1 & -1 & +1 & +1 & +1 & -1\\
+1 & -1 & -1 & +1 & +1 & -1 & -1 & +1\\
+1 & -1 & +1 & -1 & -1 & +1 & -1 & +1\\
+1 & -1 & +1 & +1 & -1 &- 1 & +1 & -1\\
+1 & +1 & -1 & -1 & -1 & -1 & +1 & +1\\
+1 & +1 & -1 & +1 & -1 & +1 & -1 & -1\\
+1 & +1 & +1 & -1 & +1 & -1 & -1 & -1\\
+1 & +1 & +1 & +1 & +1 & +1 & +1 & +1\\
\end{pmatrix*}
.
\end{align*}
\def\spacingset#1{\renewcommand{\baselinestretch}
{#1}\small\normalsize} \spacingset{1.5}
Now consider taking the difference between $\bar{Y}(\bm{z}_8) - \bar{Y}(\bm{z}_1) = \bar{Y}(+1, +1, +1) - \bar{Y}(-1, -1, -1)$, which yields
\begin{align*}
\bar{Y}(\bm{z}_8) - \bar{Y}(\bm{z}_1)&= \frac{1}{2}(\bm{h}_8-\bm{h}_1)\bm{\tau}\\
&=\tau(1)+\tau(2) + \tau(3)+\tau(123).
\end{align*}
Hence, testing whether the difference between $\bar{Y}(\bm{z}_8)$ and $\bar{Y}(\bm{z}_1)$ is zero is the same as testing whether $\tau(1)+\tau(2) + \tau(3)+\tau(123)$ is zero.
If it is reasonable to assume that all main effects are of the same sign based on subject matter knowledge and that the three-factor interaction is also of the same sign or negligible, then this global test of whether there are any treatment effects is relevant to the estimand of interest.
According to the effect hierarchy principle \citep{WuHamada}, the main effects should dominate the three-factor interaction.
Thus, even if the interaction differs in sign, we would expect to see an effect under this assumption if there is one. 
If the global test is not rejected, then we would move on to recreating the entire factorial design.
If it is unclear whether the signs of the effects are the same (also referred as antagonistic effects), then this global test would be inappropriate because effects could still be different from zero but their sum could be (close to) zero.

If we are particularly interested in the causal effects involving the first factor, we can create a test specifically for those effects.
In the the $2^3$ setting, we might use the estimand $\bar{Y}(\bm{z}_8) -\bar{Y}(\bm{z}_4)$ as a proxy for the effect of the first factor, as follows:

\begin{align}
\bar{Y}(\bm{z}_8) -\bar{Y}(\bm{z}_4)&=\bar{Y}(+1,+1,+1) -\bar{Y}(-1,+1,+1)\nonumber\\
&=\frac{1}{2}(\bm{h}_8 - \bm{h}_4)\bm{\tau}\nonumber\\
&=\tau(1)+\tau(12)+\tau(13)+\tau(123).\label{eq:y8y4}
\end{align}

In Equation~\ref{eq:y8y4}, if all terms have the same sign and we find that the difference is significantly different than zero, then we would conclude that factor 1 has an effect, either on its own or through interactions with the other factors. 
A different choice of levels for the other factors, for instance comparing the mean potential outcome when one factor is high vs. low when all other factors are at the low level, would result in different signs for the interactions.
The choice of which levels to compare should be based on subject matter knowledge and reasonable assumptions.

To reiterate, throughout this section we focused on the meaning of several estimands, which are easily estimated under a randomized experiment.
In an observational study, we would first need to obtain balance for any treatment groups we would be using in our estimator.

\section{Variance derivations}

\subsection{Variance and covariance for fractional  factorial design}\label{apped:subsecvarfracfac}
This section gives details on the derivation of the variances and covariances of our estimators of factorial effects under a fractional factorial design, as given in Equation~\ref{eq:varfracfac}.
The proofs are direct from Theorem 3 in \cite{li2017general} as we can express our estimands and estimators as linear combinations of potential outcomes.
In particular, let $\bm{A}_{\bm{z}^*_{j}}$ be the vector with entries being the $j$th entries in the $\bm{g}^*$ for all factorial effects: 
\[\bm{A}_{\bm{z}^*_{j}} = \frac{1}{2^{K-p-1}}\left([\bm{g}^*_{1}]_{[j]}, \dots, [\bm{g}^*_{K}]_{[j]}, [\bm{g}^*_{1, 2}]_{[j]}, \dots, [\bm{g}^*_{K-1, K}]_{[j]}, \dots, [\bm{g}^*_{1, 2, \dots, K}]_{[j]}\right)^T.\]
Then $\left(\tau(1), \dots, \tau(K), \tau(1,2), \dots, \tau(1,2,\dots,K)\right)^T = \sum_{j=1}^{J'}\bm{A}_{\bm{z}^*_{j}}\bm{Y}(\bm{z}^*_j)$

Then by Theorem 3 of \cite{li2017general},
\begin{align*}
\text{Var}\left(\widehat{\tau}^*(k)\right)
&= \sum_{j=1}^{J'}\frac{1}{n_j^*}[\bm{A}_{\bm{z}^*_{j}}]_{[k]}S^2(\bm{z}_j^{*})[\bm{A}_{\bm{z}^*_{j}}]_{[k]} -\frac{1}{n}\tilde{S}_k^2\\
&= \frac{1}{2^{2(K-p-1)}}\sum_{j=1}^{J'}\frac{1}{n_j^*}[\bm{g}^*_{k}]_{[j]}^2S^2(\bm{z}_j^{*}) -\frac{1}{n}\tilde{S}_k^2\\
&= \frac{1}{2^{2(K-p-1)}}\sum_{j=1}^{J'}\frac{1}{n_j^*}S^2(\bm{z}_j^{*}) -\frac{1}{n}\tilde{S}_k^2
\end{align*}

Similarly, by Theorem 3 of \cite{li2017general}
\begin{align*}
\text{Cov}\left(\widehat{\tau}^*(k), \widehat{\tau}^*(k')\right)
&= \left[\sum_{j=1}^{J'}\frac{1}{n_j^*}[\bm{A}_{\bm{z}^*_{j}}]_{[k]}S^2(\bm{z}_j^{*})[\bm{A}_{\bm{z}^*_{j}}]_{[k']} \right]-\frac{1}{n}\tilde{S}_{k,k'}^2\\
&= \frac{1}{2^{2(K-p-1)}}\left[\sum_{j=1}^{J'}\frac{1}{n_j^*}[\bm{g}^*_{k}]_{[j]}[\bm{g}^*_{k'}]_{[j]}S^2(\bm{z}_j^{*}) \right]-\frac{1}{n}\tilde{S}_{k,k'}^2\\
&= \frac{1}{2^{2(K-p-1)}}\left[\sum_{j: g_{kj}^*=g_{k'j}^*}\frac{1}{n_j^*}S^2(\bm{z}_j^*) -\sum_{j: g_{kj}^*\neq g_{k'j}^*}\frac{1}{n_j^*}S^2(\bm{z}_j^*) \right]-\frac{1}{n}\tilde{S}_{k,k'}^2
\end{align*}

\section{Relating linear regression estimators to Neyman estimators in the fractional factorial design}\label{append:linreg}
This section gives a brief overview of a proof that the linear regression point estimates are the same as the Neymanian estimates for the fractional factorial design.
For proofs of these results for the full factorial design, see \cite{lu2016randomization}.
The linear regression coefficient estimate is
\[\hat{\bm{\beta}}=(\bm{X}^T\bm{X})^{-1}\bm{X}^T\bm{Y}^{obs},\]
where $\bm{X}$ is a $n \times 2^{K-p}$ matrix whose columns correspond to first an intercept, then include the main effects, then the second order interactions not aliased with the main effects, and so on such that no two columns are aliased. 
For the $2^{3-1}$ design laid out in Section~\ref{sec:frac_fac}, $\bm{X}$ would be a design matrix with a first column of 1's and the rest of the columns corresponding to levels of the first, second,  and third factor.
No interactions would be included in this example because each of the two-factor interactions is aliased with a main effect and the three-factor interaction is aliased with the intercept, which are all already included in the model.
Thus the design looks like the columns $\bm{g}^*_1$, $\bm{g}^*_2$, and $\bm{g}^*_3$ of Table~\ref{tab:frac_fac_design_example}, but with repeated rows for each of the units assigned to the same treatment combination. 
We need $(\bm{X}^T\bm{X})^{-1}\bm{X}^T\bm{X} = \bm{I}$.
Denote $\bm{B} = (\bm{X}^T\bm{X})^{-1}\bm{X}^T$.

\[(\bm{B}\bm{X})_{ij} = \bm{b}_{i\cdot}\bm{f}_{j}\]
where $\bm{b}_{i\cdot}$ is the $i$th row of $\bm{B}$ and $\bm{f}_{j}$ is the $j$th column of $\bm{X}^{T}$.
Using notation from Section~\ref{subsec:test_sig_eff}, let $\bm{h}^*_j$ be the $j$th row of $\bm{G}^{*T}$ which is an expanded version of $\bm{z}^*_j$ which includes elements for interactions and the intercept.
Then $f_{ji}$, the $i$th element of $\bm{f}_j$ is $h^*_{kj}$, the $j$th element of $\bm{h}^*_k$, where $k$ is the treatment combination for the $i$th individual. 
It must be true that $(\bm{B}\bm{X})_{ii}=1$ and $(\bm{B}\bm{X})_{ij}=0$ for $i \neq j$.
Consider letting the $i$th row of $\bm{B}$ be $\bm{b}_{i\cdot} = \frac{1}{2^{K-p}}(\bm{\tilde{n}}^{-1}\circ\bm{f}_{i})^T$ where $\bm{\tilde{n}}^{-1}$ is the vector whose $i$th entry is $\frac{1}{n_{j}^*}$ where $j$ is number of the treatment combination that the $i$th unit is assigned to ($j = \sum_{k=1}^{J'}kW_i(z_k)$).
Then we have 
\begin{align*}
(\bm{B}\bm{X})_{kk} &= \bm{b}_{k\cdot}\bm{f}_{k}\\
&= \frac{1}{2^{K-p}}(\bm{\tilde{n}}^{-1}\circ\bm{f}_{k})^T\bm{f}_{k}\\
&= \frac{1}{2^{K-p}}(\bm{\tilde{n}}^{-1})^T\bm{f}_{k}\circ\bm{f}_{k}\\
&= \frac{1}{2^{K-p}}\sum_{j=1}^{J'}\sum_{i:W_i(\bm{z}^*_j) = 1}\frac{1}{n_j^*}\\
&= \frac{1}{J'}\sum_j1\\
&=1.
\end{align*}

For $k \neq j$
\begin{align*}
(\bm{B}\bm{X})_{kj} &= \bm{b}_{k\cdot}\bm{f}_{j}\\
&= \frac{1}{2^{K-p}}(\bm{\tilde{n}}^{-1}\circ\bm{f}_{k})^T\bm{f}_{j}\\
&= \frac{1}{2^{K-p}}(\bm{\tilde{n}}^{-1})^T\bm{f}_{k}\circ \bm{f}_{j}\\
&= \frac{1}{2^{K-p}}\sum_{s=1}^{J'}\frac{1}{n_s^*}\sum_{i:W_i(\bm{h}_s^*)=1}h^*_{sk}h^*_{sj}\\
&= \frac{1}{2^{K-p}}\left[\sum_{s:h^*_{sk}=h^*_{sj}}1-\sum_{s:h^*_{sk}\neq h^*_{sj}}1\right]\\
&=0.
\end{align*}

So the $k$th row of $\bm{B}$ is $\frac{1}{2^{K-p}}(\bm{\tilde{n}}^{-1}\circ\bm{f}_{k})^T$.
This means that
\begin{align*}
\hat{\beta}_k &= \left((\bm{X}^T\bm{X})^{-1}\bm{X}^T\bm{Y}^{obs}\right)_k\\
&=(\frac{1}{2^{K-p}}\bm{\tilde{n}}^{-1}\circ\bm{f}_{k})^T\bm{Y}^{obs}\\
&=\frac{1}{2^{K-p}}\sum_{j=1}^{J'}\frac{1}{n_j^*}\sum_{i: W_i(\bm{z}^*_j)=1}h^*_{jk}Y_i(\bm{h}^*_j)\\
&=\frac{1}{2^{K-p}}\sum_{j=1}^{J'}h^*_{jk}\bm{Y}^{obs}(\bm{z}^*_j)\\
&=\frac{1}{2^{K-p}}\bm{g}_k^{*T}\bar{\bm{Y}}^{obs}\\
&=\frac{1}{2}\widehat{\tau}^*(k).
\end{align*}
So, indeed, the linear regression estimates are one half of the factorial effects.

Now let us consider the HC2 variance estimator.
It has the form \citep{mackinnon1985some}
\[\widehat{Var}_{HC2}(\widehat{\bm{\beta}}) = \left(\bm{X}^T\bm{X}\right)^{-1}\bm{X}^T \widehat{\Omega}\bm{X}\left(\bm{X}^T\bm{X}\right)^{-1}\]
where $\widehat{\Omega} = \text{diag}\left(\frac{\hat{e}_i}{1-h_{ii}}\right)$ with $\hat{e}_i$ being the residual for observation $i$ and $h_{ii}$ being the $ii$ value of the hat matrix, $\bm{X}\left(\bm{X}^T\bm{X}\right)^{-1}\bm{X}^T$.
For discussion of this estimators in the single treatment case and the factorial case, see \citet{samii2012equivalencies} and \citet{lu2016randomization}, respectively.
We use similar ideas to those papers in the arguments below.

If unit $i$ was assigned to treatment $\bm{z}^*_k$ then the $i$th column of $\left(\bm{X}^T\bm{X}\right)^{-1}\bm{X}^T$ is $\bm{b}_{\cdot i} = \frac{1}{2^{K-p}}\frac{1}{n^*_k}\bm{h}^{*T}_k$.
We have
\begin{align*}
\left(\bm{X}\left(\bm{X}^T\bm{X}\right)^{-1}\bm{X}^T\right)_{ii} &=\frac{1}{2^{K-p}}\frac{1}{n^*_k} \bm{h}^*_k\bm{h}^{*T}_k\\
&= \frac{1}{2^{K-p}}\frac{1}{n^*_k} 2^{K-p}\\
&=\frac{1}{n_k^*}.
\end{align*}
So then $1-h_{ii} = 1-\frac{1}{n_k^*} = \frac{n_k^*-1}{n_k^*}$.
This in turn means that 
\[\frac{\hat{e}_i}{1-h_{ii}} = \frac{n_k^*\left(Y_i^{obs} - \bar{Y}^{obs}(\bm{z}_k^*)\right)^2}{n_k^*-1}\]

Now we can solve for the whole expression of $\widehat{Var}_{HC2}(\widehat{\bm{\beta}})$.
We focus on the diagonal entries.
\begin{align*}
\left(\widehat{Var}_{HC2}(\widehat{\bm{\beta}})\right)_{kk}& = \left(\left(\bm{X}^T\bm{X}\right)^{-1}\bm{X}^T \text{diag}\left(\frac{\hat{e}_j}{1-h_{jj}}\right)\right)_{i \cdot}\left(\bm{X}\left(\bm{X}^T\bm{X}\right)^{-1}\right)_{\cdot i}\\
& = (\frac{1}{2^{K-p}}\bm{\tilde{n}}^{-1}\circ\bm{f}_{k})^T \text{diag}\left(\frac{\hat{e}_j}{1-h_{jj}}\right)(\frac{1}{2^{K-p}}\bm{\tilde{n}}^{-1}\circ\bm{f}_{k})\\
& = (\frac{1}{2^{K-p}}\bm{s}_1\circ\bm{f}_{k})^T (\frac{1}{2^{K-p}}\bm{\tilde{n}}^{-1}\circ\bm{f}_{k})\\
& = (\frac{1}{2^{K-p}}\bm{s}_1)^T (\frac{1}{2^{K-p}}\bm{\tilde{n}}^{-1}\circ\bm{f}_{k}\circ\bm{f}_{k})\\
& = \frac{1}{2^{2(K-p)}}(\bm{s}_1)^T \bm{\tilde{n}}^{-1}\\
&=\frac{1}{2^{2(K-p)}} \sum_{j=1}^{J'}\frac{1}{n_j^*}\sum_{i: W_i(\bm{z}_j^*)=1}\frac{\left(Y_i(\bm{z}_j^*)-Y_i(\bm{z}_j^*)\right)^2}{n_j^*-1}\\
&=\frac{1}{2^{2(K-p)}}\sum_j\frac{1}{n_j^*}s^2(\bm{z}^*_j)
\end{align*}
where $\bm{s}_1$ is the vector of whose $i$th entry, given entry $i$ is assigned treatment combination $\bm{z}^*_k$, is $\frac{\left(Y_i(\bm{z}_k^*)-Y_i(\bm{z}_k^*)\right)^2}{n_k^*-1}$.
Thus, we have that $\left(\widehat{Var}_{HC2}(\widehat{\bm{\beta}})\right)_{kk}$ is $1/4$ times the Neyman style variance estimator.

\section{Incomplete factorial designs}\label{append:partial_des}
\subsection{Inference}
This section discusses alternative incomplete factorial designs that one might use.
See \citet{byar1993incomplete} for more discussion of incomplete factorial designs.
This discussion also applies more generally to recreating fractional designs in observational settings where we are using a subset of treatment combinations present in the data. 
We follow the same outline as proofs from Section~\ref{apped:subsecvarfracfac}.
In these incomplete factorial designs, for each main effect or other estimand of interest, we define a new experimental design tailored to estimating that effect.
When this is done, we then analyze the data as if the treatment groups within the new design are the only possible treatment groups.

First we need to introduce some notation.
Let $\dot{\bm{g}}_j$ be the same as $\bm{g}_j$ but with zero elements corresponding to treatment combinations that are not included in the particular design in use.
Let $2^m$ treatment groups be used in the design, with half assigned to the $+1$ level of factor one and half assigned to the $-1$ level of factor one.
Let $\dot{\tau}(k) = 2^{-(m-1)}\dot{\bm{g}}^{T}_k\bar{\bm{Y}}$.
We can write this using matrix form to apply Theorem 3 of \cite{li2017general}.
Define \[\dot{\bm{A}}_{\bm{z}_{j}}{} = \frac{1}{2^{m-1}}\left([\dot{\bm{g}}_{1}]_{[j]}, \dots, [\dot{\bm{g}}_{K}]_{[j]}, [\dot{\bm{g}}_{1, 2}]_{[j]}, \dots, [\dot{\bm{g}}_{K-1, K}]_{[j]}, \dots, [\dot{\bm{g}}_{1, 2, \dots, K}]_{[j]}\right)^T.\]
Then $\left(\dot{\tau}(1), \dots, \dot{\tau}(K), \dot{\tau}(1,2), \dots, \dot{\tau}(1,2,\dots,K)\right)^T = \sum_{j=1}^{J}\dot{\bm{A}}_{\bm{z}_{j}}\bm{Y}(\bm{z}^*_j)$

Then we have
\begin{align*}
\text{Var}\left(\widehat{\dot{\tau}}(k)\right) 
&=\sum_{j=1}^{J'}\frac{1}{n_j}[\dot{\bm{A}}_{\bm{z}_{j}}]_{[k]}S^2(\bm{z}_j)[\dot{\bm{A}}_{\bm{z}_{j}}]_{[k]} -\frac{1}{n}\dot{S}_k^2\\
&= \frac{1}{2^{2(m-1)}}\sum_{j=1}^{J}[\dot{\bm{g}}_{K}]_{[j]}^{2}\frac{1}{n_j}S^2(\bm{z}_j) -\frac{1}{n}\dot{S}_k^2.
\end{align*}

An important note is that in terms of estimation of this variance, whether we analyze the data as if the treatment levels in this design are the only possible treatment combinations or if we keep the assumption that units can be assigned to any possible treatment combinations (which aids in the interpretation and inference), the variance estimator will be the same.
This is because we can only estimate the first term in this expression which only involves the specific treatment levels in this design.

\subsection{Regression with missing levels}\label{reg_miss}
This section discusses what will result from a standard regression interacting all factors when not all treatment combinations are observed in the data.

If the dataset is missing $m$ treatment combinations, then the regression will be able to estimate the first $2^K-m$ effects (including interactions) that are not aliased and the rest will be removed due to collinearity of the matrix columns.
Then for each effect, there will be some aliasing structure imposed but the same ``design'' will not necessarily be used for each factor. 

To explore this scenario more, let's take the specific example of three factors where we only observe five of the eight treatment combinations.
For simplicity let there be one observation for each treatment combination and let the model matrix be as follows (we expect similar results to hold in more realistic scenarios given we use a saturated model which allows a separate mean for each treatment group):
\begin{align*}
    \bm{X} &= \begin{pmatrix}
    1 & -1 & -1 & \color{white}+\color{black}1 & \color{white}+\color{black}1\\
    1 & -1 & \color{white}+\color{black}1 & -1 & -1\\
    1 & \color{white}+\color{black}1 & -1 & -1 & -1\\
    1 & \color{white}+\color{black}1 & \color{white}+\color{black}1 & \color{white}+\color{black}1 & \color{white}+\color{black}1\\
    1 & -1 & -1 & -1 & \color{white}+\color{black}1\\
    \end{pmatrix}.
\end{align*}
The first column corresponds to the intercept, the second through fourth columns give the levels for the three factors, and the fifth column corresponds to the interaction between the first and second factor.
Note that the first four rows correspond to a $2^{3-1}$ design, so it is possible to recreate that design here.

We have
\begin{align*}
\left(\bm{X}^T\bm{X}\right)^{-1}\bm{X}^T&=
\begin{pmatrix}
0.25 & 0.25 & 0.25 & 0.25 & 0\\
-0.25 & -0.25 & 0.25 & 0.25 & 0\\
-0.25 & 0.25 & -0.25 & 0.25 & 0\\
0.5 & 0 & 0 & 0 & -0.5\\
-0.25 & -0.25 & -0.25 & 0.25 & 0.5\\
\end{pmatrix}.
\end{align*}
Recalling that $\hat{\bm{\beta}} = \left(\bm{X}^T\bm{X}\right)^{-1}\bm{X}^T\bm{Y}^{obs}$, the first three columns correspond to estimates we would get using the fractional factorial design using the defining relation $I = 123$.
The last two estimates, for factor 3 and the interaction between factors 1 and 2, have a different aliasing structure.
In particular, the aliasing on factor 3 is similar to aliasing structures in Section~\ref{subsec:test_sig_eff} and we can find that factor 3 will be aliased with the two-factor interactions 13 and 23 as well as the three-factor interaction. 

\section{Further comments on observational studies}\label{append:obs_stud}

As noted in the main text, we must assume that the treatment assignment is, at least hypothetically, manipulable such that all potential outcomes are well-defined.
Although our fractional factorial estimators do not use all treatment combinations, the estimands and aliasing structure depend on potential outcomes for the unobserved treatment combinations being well defined.
If this assumption on manipulation does not hold, we would need to consider a causal estimand that does not depend on unmeasurable or undefined potential outcomes.
Thus, it is important that subject-matter knowledge guides us in deciding which covariates are informative about which units have well-defined potential outcomes under all treatment combinations.
It may be that only a sub-population has potential outcomes for all treatments well-defined, similar to analyses for noncompliance in which the estimand is only well-defined for compliers.
For other groups where not all treatment combinations could theoretically be received, there may be other interesting estimands based on the well-defined potential outcomes, but those are not our focus.

\begin{remark}
Under some assumptions, the estimand based on only well-defined potential outcomes may be the same as the original.
If we have a $2^2$ experiment and there is no interaction between factor 1 and factor 2, then $\tau(1) = \bar{Y}(+1, +1) - \bar{Y}(-1, +1) = \bar{Y}(+1, -1) - \bar{Y}(-1, -1)$.
Hence, even if we only observed one level of factor 2 but both levels of factor 1, we could recover $\tau(1)$.
However, $\tau(1)$ as previously defined would not be of scientific interest if all potential outcomes are not defined.
\end{remark}

\section{Data illustration}\label{append:data_ill}
This section gives some additional descriptors for the data.

\begin{figure}[H]
\centering
\includegraphics[scale=0.58]{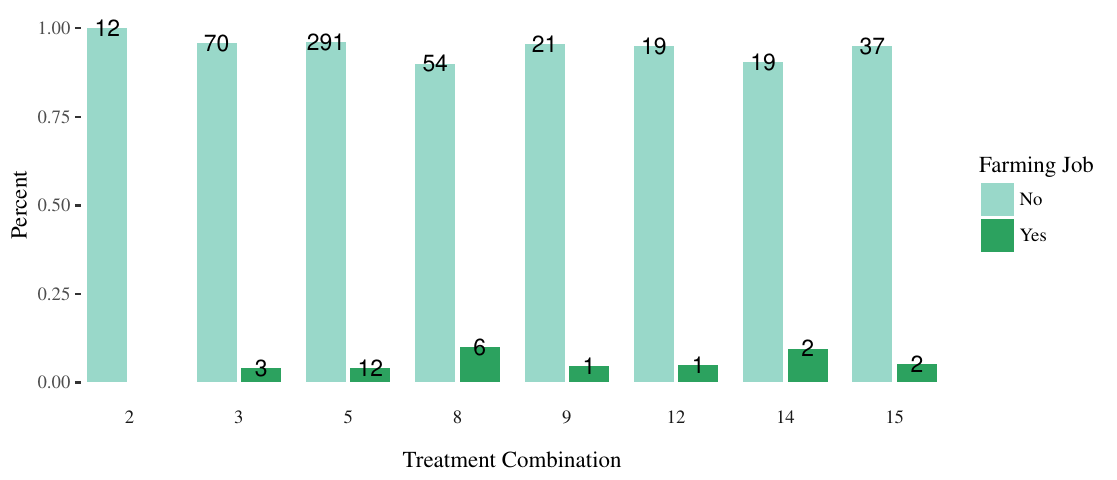}
\caption{Comparing number of farmers across factor levels in the $2^{4-1}$ fractional factorial design.
Text labels give number of observations per group.}
\label{fig:farmer_bal_bmi}
\end{figure}

Figure~\ref{fig:farmer_bal_bmi} shows that the proportion of farmers is different across the eight treatments of the hypothetical fractional factorial experiment defined in Section~\ref{sec:data_ex_design}. 

We can also examine the correlations in occurrence of the different pesticides.
beta-Hexachlorocyclohexane (beta-Hex) and heptachlor epoxide have the highest correlation of 0.48.
Mirex and heptachlor epoxide have the lowest correlation of 0.16.

For each design, we explore regression (both saturated, i.e. with all interactions among factors, and unsaturated, i.e. without interactions among factors), regression with covariates (both saturated and unsaturated for the factors), and Fisher tests for significance of effects.
We also include the HC2 standard error estimate for the saturated model without covariates, where it is equivalent to the Neyman estimator, when possible.
Note that all regression outputs are using log BMI as output.
An important note is that individuals with missing values, either for factor levels, treatment levels, or covariates where they are used, were removed.
This action likely changes the types of individuals within the analysis and therefore the generalizability of the results.
However, as this is intended as simply an illustration of the methods and not as a full analysis to draw substantive conclusions from, this simple model suffices to allow us to continue with the analysis.

Code is available for those interested in seeing the outputs from all of the analyses.
Some notes are provided for the various outputs below.

\subsection{Full factorial design}\label{append:full_fac_des}

We start by ignoring our limited data and use a full $2^4$ factorial approach.
Note that the individual who received treatment combination $(-1, 1, 1, -1)$ has leverage 1 because they are the only individual with that treatment combination.
Because of this data limitation, estimating variance using the HC2/Neyman variance estimator is not possible for the saturated case.
Also note that in the saturated model, the variance estimates given in the linear model summary are all the same.
This will be true of Neyman variance estimates too, since they are the same for each factorial effect estimators.
We see changes in significance of the estimator and even a change in sign for the estimate of beta-Hexachlorocyclohexane (beta-Hex) going from the model with just main effects to the saturated model.

Next we analyze  the full factorial design, adjusting for the covariates of income, ethnicity, gender and smoking status as linear factors in the regression model.
For simplicity, we remove all individuals who had missing values for income or ethnicity, and assume that those values are missing at random.
This resulted in 75 units being removed.
In practice one should instead use multiple imputation to account for the missing values.
One individual refused to give income (the only such individual) and so was removed.
This did not affect the analysis.
The unit assigned to the unique treatment combination, necessarily had leverage one in the saturated model.

Next we turn to the Fisherian analysis
We assume the sharp null hypothesis of zero individual factorial effects, for all factors and interactions.
This means that imputed missing potential outcomes for different assignments are just the observed potential outcomes.
We do the imputation, or effectively rearrange the assignment vector, 1000 times.
We find that only heptachlor epoxide (Hept Epox) and mirex appear to be significantly different from zero at the 0.05 level.
We only examine the main effects here but could further consider interactions.

\subsection{Fractional factorial design}\label{append:frac_fac_des}
This section gives the analysis of the fractional factorial design as is, using regression.
Note that we remove farmers again, leaving 523 observations.
The results without any covariate adjustment generally align with the full factorial analysis, in terms of sign and significance of terms.
Again it makes sense that the standard errors estimates will be the same for all estimates in the saturated model because the same groups are being used to calculate them.

Next we turn to the analysis of the fractional factorial design, adjusting for the covariates of income, ethnicity, gender and smoking status as linear factors in the regression model.
We again removed units who were missing values for income or race, reducing the sample size by 34.
One unit replied ``Don't know'' for income so this unit was removed.
This did not affect the analysis.

With the Fisherian analysis, once again, only heptachlor epoxide (Hept Ex) and mirex appear to be significantly different than zero at the 0.05 level.

\subsection{Fractional factorial with covariate adjustment}\label{append:frac_fac_mat_des}

Finally, we have the analysis of the fractional factorial design after trimming to attain balance, using regression.
169 units are retained after trimming.
We started with regressions without covariate adjustment.

\begin{figure}[!h]
\centering
\includegraphics[scale=0.6]{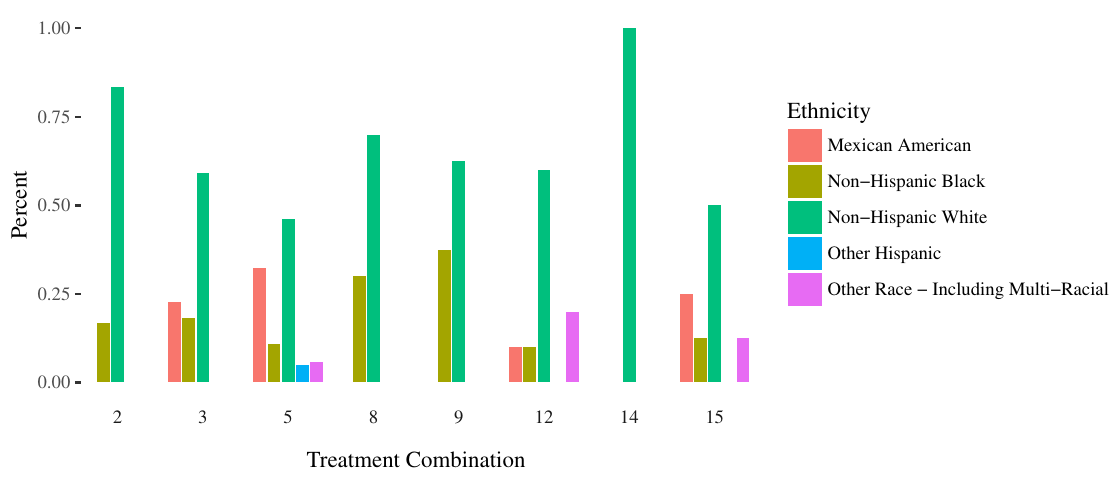}
\caption{Balance of ethnicity in the different treatment groups after matching.}
\label{fig:ethnicity_after_match}
\end{figure}

\begin{figure}
\centering
\includegraphics[scale=0.6]{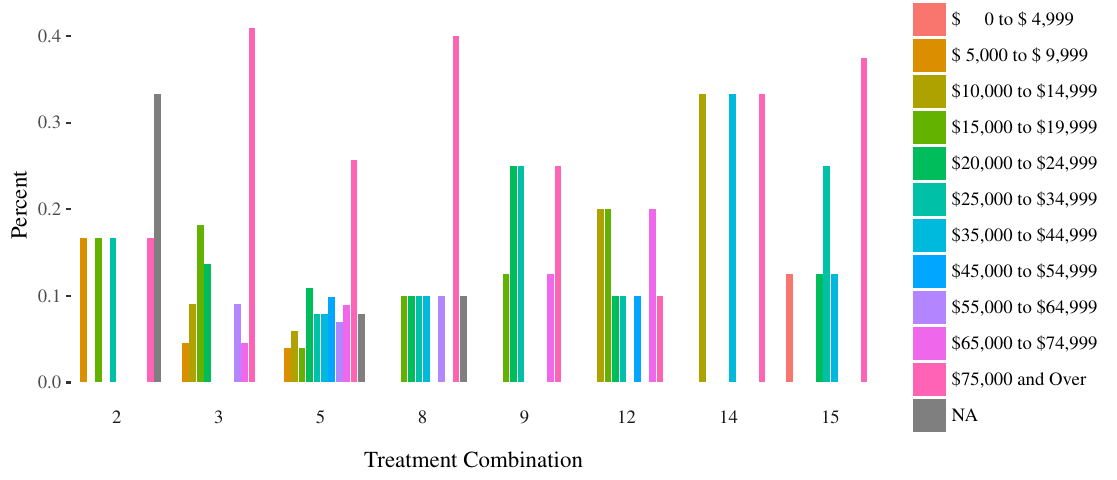}
\caption{Balance of income in the different treatment groups after matching.}
\label{fig:income_after_match}
\end{figure}

We next performed the analysis on the trimmed data set adjusting for ethnicity, income, gender, and smoking status as linear factors in the regression.
Figures~\ref{fig:ethnicity_after_match} and \ref{fig:income_after_match} shows the balance across treatment groups for income and ethnicity after trimming, indicating further need to adjust.
We see from Figure~\ref{fig:cov_bal_bmi_after_trim} that gender and smoking also require further adjustment, even after trimming.

Units who had missing values for income or ethnicity were removed, which reduced the sample size by 9.
We found that two units had unique income values of ``Over \$20,000'' and ``0 to \$4,999''.
These units were removed.
This changes the baseline level for income in the analysis from ``\$0 to \$4,999'' to ``\$5,000 to \$9,999.''

In the Fisherian analysis, mirex is the only pesticide that appears to have a significant effect on BMI at the 0.05 level.